\documentclass[11pt,tightenlines,eqsecnum,floats,aps,amsmath,amssymb,nofootinbib,prd,shownopacs,floatfix]{revtex4}
\usepackage[utf8]{inputenc}
\usepackage[T1]{fontenc}
\usepackage{amsmath,amssymb}
\usepackage{hyperref}
\usepackage{graphicx}
\usepackage{subfigure}

\usepackage{colordvi}
\usepackage{color}

\DeclareMathOperator\arctanh{arctanh}

\begin{document}
%uncomment for mobile format
% \Large
%\pacs{04.70.Dy, 02.20.Sv, 04.60.Pp}
%\title{Symmetric bounce in loop quantum Kantowski-Sachs spacetimes} 
\title{From black holes to white holes: a quantum gravitational, symmetric bounce}

\author{Javier Olmedo$^{\star,\dagger}$}
\author{Sahil Saini$^\star$}
\author{Parampreet Singh$^\star$}
\affiliation{$^\star$ Department of Physics and Astronomy,
Louisiana State University, 
Baton Rouge, LA 70803, U.S.A.}
\affiliation{$^\dagger$ Institute for Gravitation and the Cosmos, Pennsylvania State University
University Park, PA 16802, U.S.A }

\begin{abstract}
Recently a consistent non-perturbative quantization of the Schwarzschild interior resulting in a bounce from black hole to white hole geometry has been obtained by loop quantizing the Kantowski-Sachs vacuum spacetime. As in other spacetimes where the singularity is dominated by the Weyl part of the spacetime curvature, the structure of the singularity is highly anisotropic in the Kantowski-Sachs vacuum spacetime. 
As a result the bounce turns out to be in general asymmetric creating a large mass difference between the parent black hole and the child white hole. In this manuscript, we investigate under what circumstances 
a symmetric bounce scenario can be constructed in the above quantization. Using the setting of Dirac observables and geometric clocks we obtain a symmetric bounce condition which can be satisfied by a slight modification in the construction of loops over which holonomies are considered in the quantization procedure. These modifications can be viewed as quantization ambiguities, and are demonstrated in three different 
flavors which all lead to a non-singular black to white hole transition with identical masses.  Our results show that quantization ambiguities can mitigate or even qualitatively change some key features of 
physics of singularity resolution. Further, these results are potentially helpful in motivating and constructing symmetric black to white hole transition scenarios.
\end{abstract}

\maketitle

\section{Introduction}

Understanding the fate of the central singularity in the black holes is a fundamental problem whose answers lie in the quantum gravitational description of the spacetime. During the final stage of the  gravitational collapse non-perturbative quantum gravitational effects are expected to become important changing the classical singular description drastically. Though a full knowledge of the quantum gravitational effects 
modifying the dynamical collapse of an astrophysical object with an arbitrary inhomogeneous configuration is not yet available, progress in the quantization of symmetry reduced spacetimes has nevertheless provided important insights on the resolution of the central singularity. 
In particular, using techniques of loop quantum gravity Schwarzschild and other spherically symmetric spacetimes can be quantized non-perturbatively resulting in a non-singular physical description \cite{ab-ks,modesto-bh,bs-ham,husain-winkler,cartin-khanna-ks,bv,cgp-2008,gp-BH-1,gp-BH-2,chiouBH,cs-ks}. The resolution of singularity in loop quantized spacetimes is a direct manifestation of the underlying quantum geometry encoded in the spacetime constraints, which unlike differential equations in general relativity (GR) turn out to be difference equations. The latter originate from expressing the field strength of the Ashtekar-Barbero connection in terms of holonomies over loops, and provide concrete details of the dynamics of the quantum black hole spacetimes from the very small spacetime curvature scales till the Planck scale.

In these studies, Schwarzschild spacetime is the one which has been studied most rigorously. The interior of the Schwarzschild black hole is isometric to the Kantowski-Sachs metric in absence of any matter which is an anisotropic spacetime. The methods used successfully for loop quantization of cosmological spacetimes can be readily employed \cite{liv-rev,as-status,as-rev}. Using techniques similar to the ones used in loop quantum cosmology (LQC), a loop quantization of the Schwarzschild interior was first performed by Ashtekar and Bojowald \cite{ab-ks}, and by Modesto \cite{modesto-bh}. In these works a polymeric representation for the gravitational sector was introduced for the first time and the kinematical structure was studied. The quantization of this model, assuming a representation in terms of (quasi)periodic functions of the connections and following the Dirac program, has been partially analyzed in Ref. \cite{cgp-2008} and its effective dynamics deduced by means of Hamilton-Jacobi theory. The stability properties of 
the difference equation were studied in Refs. \cite{latt-ref,latt-ref-num}. Though in these works a non-singular quantum Hamiltonian constraint was obtained, the quantization suffered from problems with dependence on a fiducial length scale introduced to define the symplectic structure. This dependence resulted in unphysical and ill-defined ``Planck scale'' effects. The problem is reminiscent of the similar issues plaguing old LQC \cite{abl,aps2}. Motivated by the solution in LQC using the improved dynamics \cite{aps3}, Boehmer and Vandersloot proposed a new prescription to quantize the Schwarzschild interior \cite{bv}. While the quantization does not suffer from the problems concerning the underlying fiducial scale, it results in large quantum gravitational effects near the horizon. Recently, Corichi and Singh proposed a new quantization of the Schwarzschild interior which is free of the fiducial length scale, yields non-singular evolution and results in GR when the spacetime curvature becomes small \cite{cs-ks}. The central singularity in Corichi-Singh quantization is replaced by a quantum bounce which connects a parent black hole geometry with a child white hole geometry. In contrast to the earlier loop quantization of the Schwarzschild interior \cite{ab-ks,modesto-bh,cgp-2008}, the mass of the white hole does not depend on a fiducial length scale. The picture of the fate of the singularity resolution is also quite different from the Boehmer-Vandersloot quantization \cite{bv,chiou-int} where the singularity resolution results in a spacetime which is a product of constant curvature spaces mimicing a charged Nariai spacetime \cite{ks-constant}. Despite this difference both quantizations yield expected Planck scale effects near the central singularity with expansion and shear scalars bounded \cite{cs-ks,ks-bound,ks-strong}, a feature shared with other spacetimes in LQC \cite{ps09,ps11,cs-geom,pswe}.

A peculiar feature of Corichi-Singh quantization is the following. After the non-singular bounces of the two directional triads which capture the spatial parts of the metric, the white hole mass turns out to have a 
quartic dependence on the parent black hole mass. This disparity in the mass of the white hole seems to be universal, namely it is independent of the mass of the initial black hole, albeit within the scope of numerical investigation carried out in Ref. \cite{cs-ks}. Such a large increase in the 
white hole mass is a result of the highly asymmetric bounce in the anisotropic Kantowski-Sachs vacuum spacetime and is similar in nature to the fate of the post-bounce spacetime in other anisotropic models in LQC \cite{bgps-kasner} (for additional details about the quantum treatment of these scenarios see Refs. \cite{mmp1,mmp2,awe,mmwe}). In contrast, the quantum bounce in isotropic loop quantum cosmological spacetimes \cite{aps3,acs,mmo} is highly symmetric \cite{aps3,cs-recall,kp-fluc,cm-fluc2}. The same is assumed to be the case about the quantum gravitational regime in various phenomenological models of gravitational collapse (see Refs. \cite{bgms,gjs,tavakoli}) and on black hole to white hole transition (see for instance Refs. \cite{barcelo0,barcelo1,barcelo2,planckstar1,planckstar2}). A pertinent question is whether a symmetric bounce is possible in Corichi-Singh quantization of the Schwarzschild interior.

To gain insights on the answer to this question let us consider an important input from loop quantum gravity. Since the relationship between the loop 
quantization of symmetry reduced models and loop quantum gravity is not yet clear, it is quite possible that the symmetry reduced sector in loop quantum gravity might result in a modification of the physics 
obtained from loop quantizing the classical symmetry reduced spacetime. Evidence for this has been recently found for the case of the cosmological spacetimes where using the coherent state techniques the corresponding quantum Hamiltonian constraint turns out to  have a form which can be related to the one in LQC with a quantization ambiguity \cite{liegener}. The quantization ambiguity can be interpreted as the one in assigning different area to the loop over which holonomies are computed. This result provides support to understand the way physics of loop quantized symmetry reduced spacetimes --here Kantwoski-Sachs vacuum spacetime-- responds to modifications in the quantization procedure which 
can be attributed to quantization ambiguities. In the present context it is conceivable that such quantization ambiguities may result in a symmetric bounce scenario which is assumed in several phenomenological studies. 

The main objective of our analysis is to investigate whether a symmetric bounce is possible in the loop quantization of the Schwarzschild interior as proposed in Ref. \cite{cs-ks} or with a slight modification of the quantization. To answer this question we note that the  present model, described within either GR or the effective dynamics of loop quantum cosmology considered here, is characterized by a dynamics admitting solutions in closed form. In other words, this model is integrable (or explicitly solvable). But let us here recall that in general totally constrained theories, unfortunately, we lack an external absolute time such that a global and well-defined evolution can be constructed. Instead, the evolution is not absolute but relational. This means that one must choose a phase space function as time variable (and/or spatial coordinates in presence of spatial diffeomorphism constraints). This idea of physical clocks  was discussed in the sense of evolving constants of the motion by Rovelli \cite{carlo0,carlo1}. There, one defines the evolution by means of parametrized observables: uniparametric families of Dirac observables providing the notion of time evolution. These ideas are of special interest in the context of the quantization of totally constrained theories, given the role played by Dirac observables in order to construct the physical Hilbert space \cite{rendall0,rendall1}, like general relativity. In this context, further developments have been carried out in order to identify classical (partial) Dirac observables \cite{bianca0,bianca1} and various applications have been studied \cite{giesel-ltb,Tambornino-rev,pawlowski-dust,giesel-thiemann-scalar}.

With this in mind, the strategy that we will adopt in our analysis, as starting point, is to identify (weak) Dirac observables (constants of the motion on shell) and their conjugate momenta. This serves two purposes. On the one hand, to identify a condition for symmetric bounce in terms of geometrical clocks, and, on the other hand, to lay the platform for a subsequent reduced phase space quantization of the model. A primary step in our analysis is to carry out a canonical transformation from the old Ashtekar-Barbero variables to the Dirac observables. We then proceed with the study of the dynamics. For the latter one can either choose a lapse function and then solve the equations of motion or, equivalently, one can implement a suitable gauge-fixing condition (second-class constraint) in this gauge system and solve the dynamics of the reduced (or true) Hamiltonian. In general, the conjugate variables to the weak Dirac observables provide a natural internal geometrical clock (or physical time), since they are usually well-behaved monotonic functions (up to topological obstructions). Although we will not study here the dynamics of the model quantum mechanically, our analysis of the classical and effective (loop quantum cosmology) dynamics will definitely be very useful in future analyses in the context of reduced phase space and full quantizations, either within a standard or a polymeric representation.

Our manuscript is organized as follows. In Sec. II we begin with an outline of the classical Hamiltonian constraint in terms of symmetry-reduced connection and triad variables and after identifying two (weak) Dirac observables we rewrite the Hamiltonian constraint in terms of the latter. Then we perform a gauge fixing which identifies an internal clock. (An alternative gauge fixing is discussed in Appendix A). In Sec. III this exercise is repeated for the effective Hamiltonian constraint of the loop quantization of the Schwarzschild interior based on the analysis of Ref. \cite{cs-ks}. The Dirac observables and the internal clock identified in the effective spacetime description yield the ones in the classical theory when the quantum discreteness vanishes. (An alternative to the gauge fixing used in Sec. III is discussed in Appendix B). In Sec. IV we identify the condition to obtain the symmetric bounce in the black hole to white hole transition. We show that this condition can not be satisfied for any real mass for Corichi-Singh quantization, but can be satisfied if modifications are made to the assignment of the minimum area of the loop over which holonomies are considered. Three such choices of modifications are considered, along with choice 0 corresponding to 
Corichi-Singh construction. These choices are parameterized through two parameters $\alpha$ and $\beta$ whose values can be fixed given the initial black hole mass. We then discuss numerical results from various choices which provide insights on so far not known phenomenological features of black hole to white hole transition. These include minimum allowed masses, existence of quantum recollapse in highly quantum black holes and non-linear behavior of the value of the volume at the bounce with respect to black hole mass. We summarize our main results in Sec. V.

\section{Classical setting}

We start with a discussion of some of the main aspects of the classical theory of the model  considered in this manuscript. In order to describe the interior of a black hole in real Ashtekar-Barbero variables we adopt a symmetry reduction \cite{ab-ks}, such that the spatial slices are compatible with the Kantowski--Sachs symmetry group $G = \mathbb{R} \times SO(3)$.\footnote{There, the invariant connection and triad lead to a vanishing diffeomorphism constraint, while the Gauss constraint remains nontrivial. However, we impose this constraint by first identifying suitable phase space functions that commute with it, and either carrying out a canonical transformation that splits the phase space between gauge invariant and pure gauge variables or by introducing a gauge fixing \cite{ab-ks}.} Besides, we will consider a recent treatment carried out in Ref. \cite{cs-ks}, where a careful implementation of fiducial structures is taken into account. The connection and the densitized triad take the form
\begin{equation}
	A^i_a \, \tau_i \, {\rm d} x^a \, = \, \bar c \, \tau_3 \, {\rm d} x + \bar b\,r_o \, \tau_2 {\rm d} \theta 
	- \bar b \,r_o\, \tau_1 \sin \theta \, {\rm d} \phi + \tau_3 \cos \theta \, {\rm d} \phi,
\end{equation}
and
\begin{equation}
	E^a_i \, \tau^i \frac{\partial}{\partial x^a} \, =  \, \bar p_c \,r_o^2\, \tau_3 \, \sin \theta 
	\, \frac{\partial}{\partial x} + \bar p_b\, r_o \, \tau_2 \, \sin \theta \, 
	\frac{\partial}{\partial \theta} - \bar p_b\, r_o \, \tau_1 \,  \frac{\partial}{\partial \phi},
\end{equation}
where $x\in [0,L_o]$, $\theta$ and $\phi$ are the typical angular coordinates of the two-sphere of unit radius.  Besides, $\tau_i$ are the standard basis elements of $su(2)$ fulfilling $[\tau_i,\tau_j]=\epsilon_{ijk}\tau^k$. Finally, $r_o=2GM$ (Schwarzschild radius), and $(\bar b, \bar p_b)$ and $(\bar c, \bar p_c)$ are the coordinates of the reduced phase space of this system.

The spacetime metric in terms of the triads is given by
\begin{equation}
	{\rm d} s^2 = - N^2 {\rm d} t^2 + \frac{\bar p_b^2}{|\bar p_c|} \, {\rm d} x^2 + |\bar p_c|\,r_o^2 \, 
	({\rm d} \theta^2 + \sin^2 \theta\, {\rm d} \phi^2).
\end{equation}
We then choose the triad components $\bar p_b$ and $\bar p_c$ to be dimensionless, and such that $\bar p_c$ is equal to unity at the horizon. The canonical Poisson brackets, respectively, are
 \begin{equation}\label{old-pbs}
	\{\bar c,\bar p_c\} \, = \frac{\,2 G \gamma}{L_or_o^2}, ~~~~ \{\bar b,\bar p_b\} \, = \,  \frac{G \gamma}{L_or_o^2},
\end{equation}

Now, we introduce the new set of phase space variables
\begin{equation}\label{cbpair}
	c = L_o \, \bar c, ~~ p_c = r_o^2\,\bar p_c, ~~ b = r_o\,\bar b, ~~ p_b = r_o\,L_o \, \bar p_b,
\end{equation}
such that the new Poisson brackets 
\begin{equation}\label{pbs}
	\{c,p_c\} \, = \,2 G \gamma, ~~~~ \{b,p_b\} \, = \,  G \gamma,
\end{equation}
are independent of $r_o$ or $L_o$.

One can show that the total Hamiltonian of the system is 
\begin{equation}
	H_{\mathrm{class}} = \frac{1}{16 \pi G}NC_{\mathrm{class}},
\end{equation}
where $C_{\mathrm{class}}$ is the Hamiltonian constraint
\begin{equation}\label{h_cl}
	C_{\mathrm{class}} = - \frac{8 \pi \mathrm{sgn} (p_c)}{\gamma^2}\left((b^2 + \gamma^2) \frac{p_b}{\sqrt{|p_c|}} + 2 b c |p_c|^{1/2} \right).
\end{equation}
Note that the diffeomorphism constraint identically vanishes due to the symmetry reduction. 

\subsection{Canonical transformation: Dirac observables}

In this particular model, it is possible to identify a couple of phase space variables that weakly commute with the Hamiltonian constraint. They correspond to
\begin{align}\label{gr-obs-conf}
	o_1&=\frac{(\gamma^2 + b^2)p_b}{b}, \quad \mathrm{and} \quad o_2= 2 c p_c.
\end{align}
Let us mention that, on shell, they are not independent from each other. Therefore, on shell, only one of them will correspond to a weak Dirac observable. In addition, we can introduce two conjugate momenta to these variables, namely
\begin{align}\label{gr-obs-momen}
	p_1&=-\frac{1}{2 G \gamma}\log\left(1 + \frac{b^2}{\gamma^2}\right), \quad \mathrm{and} \quad  p_2= -\frac{1}{4 G \gamma}\log|c|,
\end{align}
respectively. One can see that $o_1\in(-\infty,\infty)$ and $p_1\in(-\infty,0)$ if $b\in(-\infty,\infty)$ and $p_b\in(-\infty,\infty)$. Besides, $o_2\in(-\infty,\infty)$ and $p_2\in(-\infty,\infty)$ for $c\in(-\infty,\infty)$ and $p_c\in(-\infty,\infty)$. 

These new variables satisfy the Poisson algebra
\begin{align}
\{o_i,p_j\}=\delta_{ij},\quad \{o_i,o_j\}=0,\quad \{p_i,p_j\}=0,
\end{align}
for $i,j=1,2$.

We can invert the previous relations and obtain the original phase space variables as functions of the new configuration and momenta as
\begin{align}\label{eq:c-to-Oi}
	c &= \epsilon_c e^{-4 G \gamma  p_2},\quad p_c=\frac{1}{2}\epsilon_co_2e^{4 G \gamma p_2}, \\\label{eq:b-to-Oi}
	b &= \epsilon_b \gamma e^{-G \gamma p_1} \sqrt{1 - e^{2 G \gamma p_1}}, \quad p_b=\epsilon_b\frac{1}{\gamma}o_1e^{G \gamma p_1} \sqrt{1 - e^{2 G \gamma p_1}}.
\end{align}
Here, $\epsilon_c$ and $\epsilon_b$ are equal to $\pm1$.
In terms of these observables, the Hamiltonian constraint takes the following form
\begin{equation}\label{h_cl1}
	C_{\mathrm{class}} = - \frac{8 \pi \epsilon_c\epsilon_b\,\mathrm{sgn} (o_2)}{\gamma} e^{-G \gamma (p_1+2p_2)} \sqrt{\frac{2}{|o_2|}}\sqrt{1 - e^{2 G \gamma p_1}}\left(o_1 + o_2 \right) ~.
\end{equation}

\subsection{Classical dynamics}\label{sec:class-dyn}

The classical dynamics can be solved by means of the Hamilton equations of the system. They can be obtained after computing Poisson brackets of the basic phase space variables with the Hamiltonian. However, since the total Hamiltonian is a constraint (obtained after varying the action with respect to the lapse function), not only the initial data must be specified within the constraint surface but also there is no preferred choice of time in the system. In order to integrate the equations of motion, one should specify (as a prescribed function) the lapse, which plays the role of a Lagrange multiplier. 

However, we will adopt an equivalent treatment. Concretely, we will provide below an example where, instead of solving the equations of motion after specifying the lapse function, we will introduce a suitable gauge fixing condition for one of the conjugate momenta of the weak Dirac observables defined above. This gauge fixing selects one of the phase space variables as a physical (internal) clock (see Appendix \ref{app:A} for an alternative gauge fixing). With respect to this clock, we can compute a reduced (or true) Hamiltonian ruling the dynamics of the reduced system, which can actually be easily solved.\footnote{This procedure is in agreement with the evolving constants introduced by Rovelli \cite{carlo0,carlo1}. See Refs. \cite{bianca0,bianca1} for an extended discussion.} Finally, using Eqs. \eqref{eq:c-to-Oi} and \eqref{eq:b-to-Oi}, we can explicitly write the original phase space variables as functions of Dirac observables and time. 

Let us start with the following gauge fixing condition $\Phi=p_2-\tau\approx 0$, where $\tau$ plays the role of time parameter. We will assume, a priori, that it can take any value in the real line. Preservation of this condition upon evolution, after evaluation on shell (i.e. $C_{\rm class} \approx 0$ and $\Phi\approx 0$), determines the lapse function, namely
\begin{align}
0\approx\dot\Phi=\{p_2,H_{\rm class}\}-1.
\end{align}
In terms of the new variables the lapse takes the following form (on shell)
\begin{equation}\label{lps-g1}
	N \approx \frac{\gamma}{8 \pi \epsilon_c\epsilon_b\,\mathrm{sgn} (-o_1)}\frac{e^{G \gamma (p_1+2\tau)}}{\sqrt{1 - e^{2 G \gamma p_1}}}  \sqrt{\frac{|o_1|}{2}},
\end{equation}
after employing the constraint defined in Eq. \eqref{h_cl1} and the gauge condition, namely $o_2\approx-o_1$ and $p_2\approx \tau$. Eventually, the replacement of these conditions in the full action allows us to obtain the reduced action as
\begin{equation}
	S_{\rm red} = \int d\tau \left(p_1\dot{o}_1-h_{\rm red}\right),
\end{equation}
with $h_{\rm red}=-o_1$ as the reduced Hamiltonian. The dynamics can be easily solved. Here, $o_1$ is a constant of the motion and $p_1=\tau+p_1^0$, with $p_1^0$ another constant of the motion. Now, let us notice that the definition of $p_1$ given in Eq. \eqref{gr-obs-momen} involves $p_1<0$, namely, the parameter $\tau$ must fulfill $\tau<-p_1^0$.

Finally, let us understand the physical meaning of the observables $(o_1,p_1^0)$. In order to do so, it is convenient to recall that in the interior of the black hole we have a clear interpretation regarding the values that the triads and connections take at the horizon. Concretely, if the mass of the black hole is $M$, we choose the standard conditions (see for instance Ref. \cite{cs-ks}) at the horizon $p_b(\tau_{\rm hor})=0=b(\tau_{\rm hor})$, $p_c(\tau_{\rm hor})=(2 G M)^2$ and $c(\tau_{\rm hor})=\frac{\gamma L_o}{4 G M}$. Now, if we express the original variables in terms of the new ones by means of Eqs. \eqref{eq:c-to-Oi} and \eqref{eq:b-to-Oi}, we set $o_2\approx-o_1$, $p_2\approx \tau$ and $p_1=\tau+p_1^0$, we obtain 
\begin{align}
	c(\tau) &=  \frac{\gamma L_o}{4GM}e^{-4 G \gamma  (\tau-\tau_{\rm hor})},\quad p_c(\tau)=(2GM)^2e^{4 G \gamma (\tau-\tau_{\rm hor})}, \\
	b(\tau) &= - \gamma e^{-G \gamma (\tau-\tau_{\rm hor})} \sqrt{1 - e^{2 G \gamma (\tau-\tau_{\rm hor})}}, \quad p_b(\tau)= 2GML_oe^{G \gamma (\tau-\tau_{\rm hor})} \sqrt{1 - e^{2 G \gamma (\tau-\tau_{\rm hor})}},
\end{align}
where 
\begin{equation}
	\tau_{\rm hor}=\frac{1}{4\gamma G}\log\left(\frac{4GM}{\gamma L_o}\right).
\end{equation}
Besides, one can easily see that 
\begin{equation}
o_1= -2GM\gamma L_o,\quad p_1^0= -\tau_{\rm hor}.
\end{equation}
Therefore, the Dirac observables are functions completely determined by the mass of the black hole. Let us also comment that, here, we have restricted the solutions to the sector $p_b(\tau)\geq 0$ and $p_c(\tau)\geq 0$ for all $\tau$, which involve $\epsilon_c=1$ and $\epsilon_b=-1$. Besides, the physical time fulfills $\tau\in(-\infty,\tau_{\rm hor}]$. Then, the singularity is reached at $\tau\to-\infty$.

\section{Effective Loop Quantum Cosmology}

In loop quantum cosmology, the situation changes drastically from a physical point of view, with respect to the previous classical model. The classical singularity, in this particular model, has been shown to be replaced by a quantum bounce \cite{cs-ks,cgp-2008,bv,ks-bound} within different dynamical schemes. Interestingly, the quantum bounce turns out to be nonsymmetric in general. As a consequence, if one starts the evolution of the spacetime corresponding to a semiclassical black hole of mass $M$, after the bounce, one reaches a new semiclassical geometry that in general will not agree with the one of a black (white) hole of the same mass $M$. Concretely, in Ref. \cite{cs-ks} it was shown that, in addition to a suitable treatment of the fiducial structures present in the theory, the final physical picture provides a black hole interior of mass $M$ connected through a bounce with a white hole interior of mass generally different from the one associated with the initial black hole. This scenario corresponds to a nonsymmetric bounce. 

Here, we will explore several available dynamical schemes within a family of choices closely related to the one studied in Ref. \cite{cs-ks} with the purpose of reconciling the loop quantum effective dynamics of this model with symmetric bounces, {\it regardless of the initial conditions}. 
Our starting point in order to build the effective Hamiltonian constraint is the classical setting analyzed in the previous section in real Ashtekar-Berbero variables. We must then remind that in the quantum theory, the connection is not a well defined operator. Actually, holonomies of the connection provide a suitable algebra of kinematical operators. Then, following the canonical quantization program of loop quantum gravity, we must define an operator representing the Hamiltonian constraint. It has been carried out in detail, for instance, in Refs. \cite{ab-ks,cs-ks,cgp-2008}. There, the curvature operator is written as a combination of holonomies around closed loops. In the full theory, the limit in which the area enclosed by these loops goes to zero can be taken \cite{qsd}. However, in the present homogeneous model, this limit is not well defined. Concretely, we consider loops of holonomies in the $x-\theta$, $x-\phi$ and $\theta-\phi$ planes. Edges of the holonomies along $x$ have length $\delta_c L_o$ while edges along longitudes and equator on the two-sphere have length $\delta_b r_o$. It should be noted that the loop in $\theta-\phi$ plane is not a closed loop but given the homogeneity of space it can still be assigned an `effective' area \cite{ab-ks}. 

The effective Hamiltonian of the system, in terms of holonomies along the previous edges, is\footnote{The effective Hamiltonian constraint has been studied in several Refs. \cite{ab-ks,cs-ks,cgp-2008,bv,ks-bound}. The validity of effective Hamiltonian has been tested for various isotropic and anisotropic models for a wide variety of states \cite{mop,numlsu-2,numlsu-3,numlsu-4}.}
\begin{equation}
H_{\mathrm{LQC}} = \frac{1}{16 \pi G}NC_{\mathrm{LQC}},
\end{equation}
where $C_{\mathrm{LQC}}$ is the effective Hamiltonian constraint in loop quantum cosmology and takes the form
\begin{equation}\label{h_lqc}
C_{\mathrm{LQC}} = - \frac{8 \pi \mathrm{sgn} (p_c)}{\gamma^2}\left[\left(\frac{\sin^2(\delta_b b)}{\delta_b^2} + \gamma^2\right) \frac{p_b}{\sqrt{|p_c|}} + 2 \frac{\sin(\delta_b b)}{\delta_b} \frac{\sin(\delta_c c)}{\delta_c}  |p_c|^{1/2} \right] .
\end{equation}

It is worth commenting that the dynamics generated by this effective Hamiltonian can be solved explicitly, in a similar fashion as it was done in the classical theory. Here, we will focus in a set of dynamical schemes where $\delta_b$ and $\delta_c$ are constants (they will not depend on dynamical phase space variables). Therefore, we do not need to specify the length of the edges now in order to solve the dynamics. The precise form of $\delta_b$ and $\delta_c$ will be explained in Sec. \ref{sec:dyn-prescs}.

\subsection{Canonical transformation: Dirac observables}

As in the classical theory, it is possible to identify two phase space variables that (weakly) commute with the effective Hamiltonian constraint under Poisson brackets, namely
\begin{align}
O_1&=\left(\frac{\sin(\delta_b b)}{\delta_b} +\frac{ \gamma^2\delta_b}{\sin(\delta_b b)}\right)p_b, \quad \mathrm{and} \quad O_2= 2 \frac{\sin(\delta_c c)}{\delta_c} p_c.
\end{align}
If the effective constraint \eqref{h_lqc} is imposed, these two phase space variables are not linearly independent. Hence, only one of them (or a combination of both) codifies the physical degree of freedom, i.e. only one Dirac observable. Two conjugate momenta to these variables are
\begin{align}
P_1&=\frac{1}{2 G \gamma b_o}\log\left(\frac{1 + \frac{\cos(\delta_bb)}{b_o}}{1 - \frac{\cos(\delta_bb)}{b_o}}\right), \quad \mathrm{and} \quad P_2= -\frac{1}{4 G \gamma}\log\left|\tan\left(\frac{\delta_cc}{2}\right)\right|,
\end{align}
respectively, where $b_o=\sqrt{1+\gamma^2\delta_b^2}$. One can see that $O_1\in(-\infty,\infty)$ and $P_1\in(-P_1^{\rm max},P_1^{\rm max})$ if $b\in(-\infty,\infty)$ and $p_b\in(-\infty,\infty)$, with
\begin{equation}
P_1^{\rm max}=\frac{1}{2 G \gamma b_o}\log\left(\frac{1 + \frac{1}{b_o}}{1 - \frac{1}{b_o}}\right).
\end{equation}
Besides, $O_2\in(-\infty,\infty)$ and $P_2\in(-\infty,\infty)$ for $c\in(-\infty,\infty)$ and $p_c\in(-\infty,\infty)$. 

Let us notice that the relations between $O_i$ and $(c,b)$ involve trigonometric functions. Therefore, there is some degeneracy that we fix by restricting the connections $(c,b)$ to the interval $[-\pi/\delta_c,\pi/\delta_c)$ and $[-\pi/\delta_b,\pi/\delta_b)$, respectively. 

The Poisson algebra of these new variables is
\begin{align}
\{O_i,P_j\}=\delta_{ij},\quad \{O_i,O_j\}=0,\quad \{P_i,P_j\}=0.
\end{align}

The canonical transformation can be inverted to obtain symmetry reduced connection and triad components in terms of $(O_1,P_1)$ and $(O_2,P_2)$:
\begin{align}\label{eq:cb-to-O-lqc}
c &= \frac{2}{\delta_c}\arctan\left(e^{-4 G \gamma  P_2}\right),\quad p_c=\epsilon_c\frac{\delta_c}{2}O_2\cosh\left(4 G \gamma P_2\right), \\
b &= \epsilon_b\frac{1}{\delta_b}\arccos\left[b_o\tanh\left(G \gamma b_o P_1\right)\right], \quad p_b=\epsilon_b\frac{\delta_b O_1}{b_o^2} \cosh^2\left(G \gamma P_1 b_o\right) \sqrt{1 - b_o^2 \tanh^2\left[G \gamma P_1 b_o\right]}.
\end{align}
Here, as before, $\epsilon_b$ and $\epsilon_c$ are equal to $\pm1$. 

Furthermore, the effective Hamiltonian constraint in terms of these new variables takes the form
\begin{equation}\label{h_cl1-lqc}
C_{\mathrm{LQC}} = - \epsilon_c\epsilon_b\frac{8 \pi}{\gamma^2}\frac{\frac{1}{\delta_b}\sqrt{1 - b_o^2 \tanh^2\left[G \gamma P_1 b_o\right]}}{\sqrt{\frac{\delta_c}{2}|O_2|\cosh\left(4 G \gamma P_2\right)}}\left(O_1+O_2\right) ~.
\end{equation}

\subsection{Effective dynamics}\label{sec:eff-dyn}

As in the classical theory, in order to solve the effective dynamics in LQC, we will adopt the gauge fixing $\Psi_1=P_2-T \approx 0$, where $T$ is the time parameter for this gauge fixing and it can take any value in the real line (for an alternative gauge fixing see Appendix \ref{app:B}). Preservation of this condition upon evolution, after evaluation on shell,
\begin{align}
0\approx\dot{\Psi}=\{P_2,H_{\mathrm{LQC}}\}-1,
\end{align}
fixes the lapse 
\begin{equation}\label{eq:lapse}
N=\epsilon_c\epsilon_b\frac{\gamma^2}{8 \pi}\frac{\sqrt{\frac{\delta_c}{2}|O_1|\cosh\left(4 G \gamma T\right)}}{\frac{1}{\delta_b}\sqrt{1 - b_o^2 \tanh^2\left[G \gamma b_oP_1\right]}}.
\end{equation}

Besides, the new configuration variables, on shell, satisfy $O_2=-O_1$. Eventually, one can easily see that the reduced action takes the form
\begin{equation}
S_{\rm red} = \int dT \left(P_1\dot{O}_1-H_{\rm red}\right),
\end{equation}
with $H_{\rm red}=-O_1$ the reduced Hamiltonian. As a consequence, one easily realizes that $O_1$ is actually a constant of the motion while $P_1=T+P_1^0$, with $P_1^0$ another constant of the motion. Besides, $T$ must be constrained such that $|P_1|\leq P_1^{\rm max}$, i.e. $|T+P_1^0|\leq P_1^{\rm max}$.

Eventually, in order to connect the dynamics in classical GR with the effective dynamics of LQC, we first identify the value of the Dirac observable in both descriptions. This amounts to   
\begin{equation}
O_1= -2GM\gamma L_o.
\end{equation}
Besides, we also identify the connections at the horizon, i.e. $c(T_{\rm hor})=\gamma L_o/(4GM)$ and $b(T_{\rm hor})=0$. This conditions allows to specify the time $T_{\rm hor}$ and the Dirac observable $P_1^0$. Concretely, 
\begin{equation}\label{eq:init-dt}
T_{\rm hor}=\frac{1}{4\gamma G}\log\left(\frac{8GM}{\gamma L_o\delta_c}\right),
\quad P_1^0=\frac{1}{b_o G\gamma}\arctanh\left[\frac{1}{b_o}\right]-T_{\rm hor}.
\end{equation}

Then, we can write the triads and connections as functions of $T$ and the constants of the motion $O_1$ and $P_1^0$ using Eqs. \eqref{eq:cb-to-O-lqc}. We obtain
\begin{subequations}\label{eq:cb-to-T-lqc}
\begin{align}
c(T) &= \frac{2}{\delta_c}\arctan\left(e^{-4 G \gamma  T}\right),\quad p_c(T)=-\epsilon_c\frac{\delta_c}{2}O_1\cosh\left(4 G \gamma T\right), \\
b(T) &= \epsilon_b\frac{1}{\delta_b}\arccos\left[b_o\tanh\left(G \gamma b_o (T+P_1^0)\right)\right], \\
p_b(T)&=\epsilon_b\frac{\delta_b O_1}{b_o^2} \cosh^2\left(G \gamma (T+P_1^0) b_o\right) \sqrt{1 - b_o^2 \tanh^2\left[G \gamma (T+P_1^0) b_o\right]}.
\end{align}
\end{subequations}

\section{Dynamical prescriptions for symmetric bounces}\label{sec:dyn-prescs}

Now that the effective dynamics has been solved (for constant $\delta_b$ and $\delta_c$), we can come to the main objective of this manuscript. We will prove here that this effective model admits symmetric bounces, regardless of the initial conditions (and the mass of the black hole). This condition is fulfilled if and only if complete dynamical trajectories in the phase space plane $p_b$-$p_c$ enclose a zero area. By complete trajectory we mean that they start and end at the horizons. In other words, if we start the evolution in a collapsing branch, once the trajectory reaches the bounce it must retrace the same phase space points during the expanding branch (by continuity the area enclosed will be zero) reaching the same phase space point it started at.\footnote{{This is analogous to the situation in magnetism corresponding to zero hysteresis.}} This condition is fulfilled if and only if $P_1^0=0$. Then, symmetric bounces involve 
\begin{equation}\label{eq:sym-bcond}
\frac{\arctanh\left[\frac{1}{\sqrt{1+\gamma^2\delta_b^2}}\right]}{\sqrt{1+\gamma^2\delta_b^2}}=\frac{1}{4}\log\left(\frac{8GM}{\gamma L_o\delta_c}\right).
\end{equation}
In consequence, this (symmetric bounce) condition requires that the polymer parameters $\delta_c$ and $\delta_b$ will not be independent, namely, a symmetric bounce is a physical requirement that eliminates some the freedom in the choices of $\delta_c$ and $\delta_b$.

Actually, this freedom can be traced back to the available choices for the finite length of the edges of the holonomies. For instance, in Ref. \cite{cs-ks} it was proposed to fix these parameters as
\begin{equation}\label{eq:old-poly}
\delta_b^2 r_o^2=\Delta,\quad \delta_b r_o \delta_c L_o=\Delta.
\end{equation}

We will codify this freedom in two constant parameters $\alpha$ and $\beta$ such that conditions \eqref{eq:old-poly} become
\begin{equation}\label{eq:new-poly}
\delta_b^2 r_o^2=\alpha^2\Delta,\quad \delta_b r_o \delta_c L_o=\alpha\beta\Delta.
\end{equation}
If we recall that $r_o$ is the Schwarzschild radius and that $L_o$ is a fiducial length to be specified, we must notice that together with the symmetric bounce condition \eqref{eq:sym-bcond}, one can easily see that the parameters $\alpha$ and $\beta$ will be independent of $L_o$, but they will depend on $r_o$, namely, they will depend on the mass of the black hole. Note that this kind of dependence on the mass has been already suggested in Ref. \cite{cs-ks}. The difference here is that such dependence will be more involved. In some situations it will not be possible to provide the relation in a closed form. 

Below, we will analyze some possible choices for these parameters. The reader must notice that, among the simplest choices, there is a natural parametrization corresponding to $\beta=1$. It is due to the freedom in the choice of the open loop in the $\theta-\phi$ plane. Nevertheless, we will consider other simple choices, for the sake of completeness, such as $\alpha=1$ and $\alpha=1/\beta$. Before, it will be convenient to first summarize the results found in Ref. \cite{cs-ks} where asymmetric bounces occur. 

In all the numerical computations that will be shown below, we set  $\Delta=4\sqrt{3}\pi\gamma G\hbar$, $\gamma=0.2375$, $L_o=1$, $G=1$ and $\hbar=1$. We start with the discussion of the choice made in Corichi-Singh quantization \cite{cs-ks}, followed by three choices to obtain symmetric bounces.

\subsection{Choice 0: asymmetric bounce}

\begin{figure}
	\centering
	\includegraphics[width=0.49\textwidth]{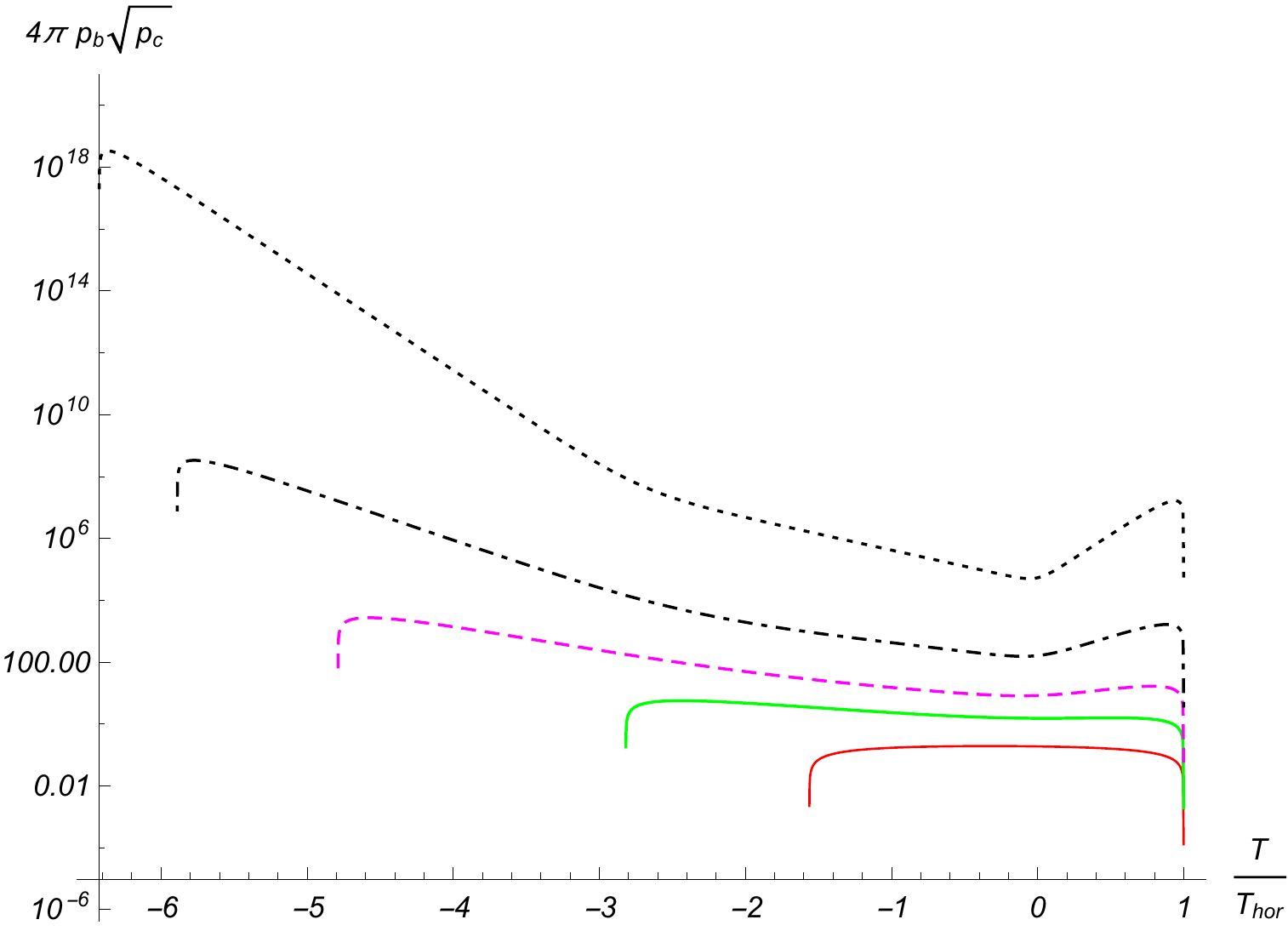} 
	\includegraphics[width=0.49\textwidth]{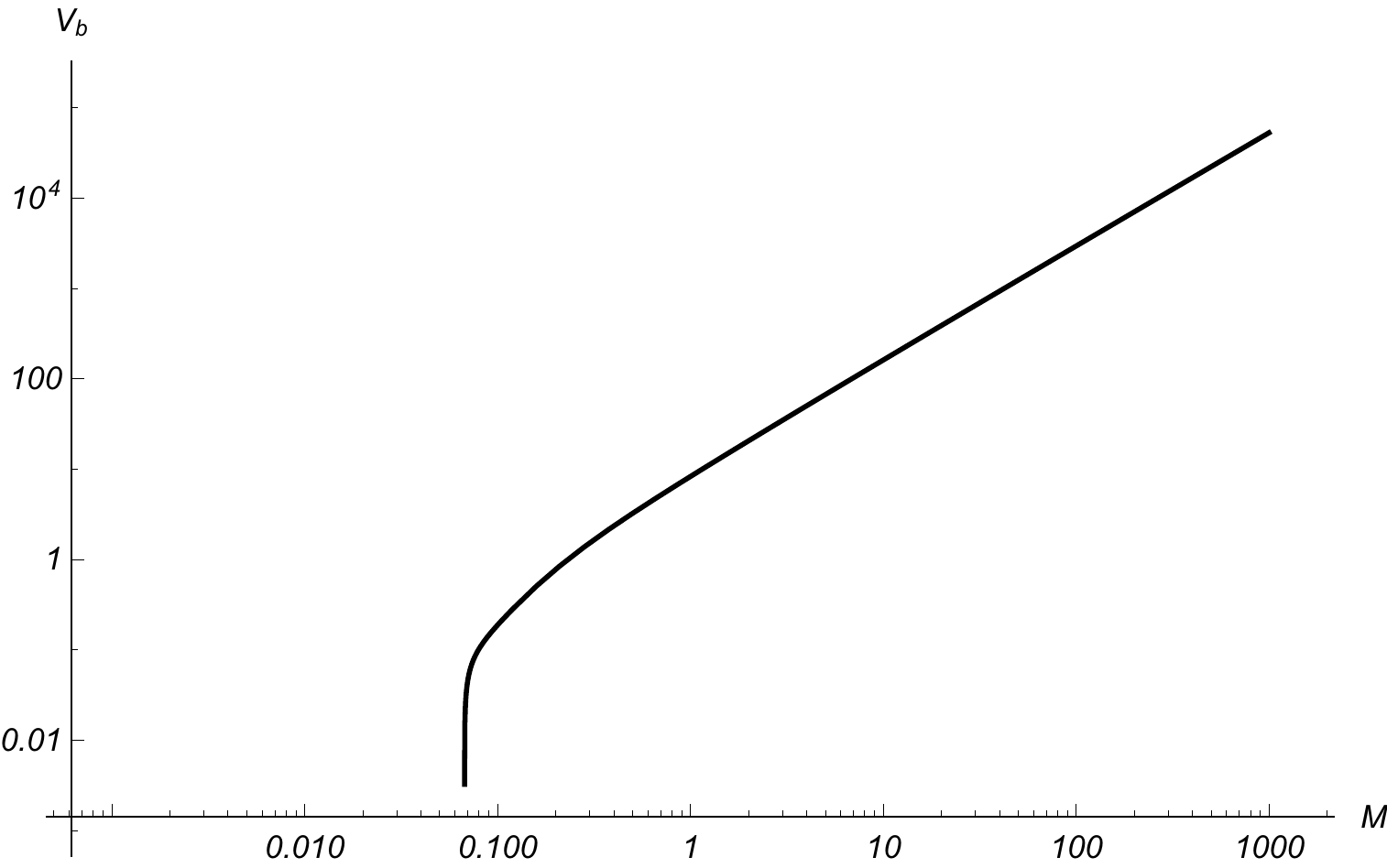} 
	\caption{The left graph provides the evolution of the spatial volume as a function of the normalized time $T/T_{hor}$. The curves ranging from thickest to the thinnest correspond to $M=0.1,0.3,1,10,1000$ (in Planck units), respectively. In the right graph we plot the values of the volume at the bounce as a function of the mass.}
	\label{fig:opt0}
\end{figure}

In this case, we reproduce the results already shown in Ref. \cite{cs-ks}. Namely, the polymer parameters $\delta_b$ and $\delta_c$ are fixed by Eq. \eqref{eq:old-poly}. We then plug them in Eq. \eqref{eq:cb-to-T-lqc} and, for this choice, we summarize our study of the dynamics in Fig. \ref{fig:opt0} using two plots. In the left one we show the spatial volume $V = 4 \pi p_b\sqrt{p_c}$ as a function of time $T$, for several values of the mass of the black hole, from semiclassical to very quantum black holes, namely, from large to small masses w.r.t the Planck mass, respectively. We see that the volume goes to zero at the horizons, since $p_b$ does. Besides, for large enough values of the mass, the volume at each side of the bounce (minimum value reached by the volume away from the horizons) is different. This is an indication that the bounce is not symmetric. However, if the masses are small enough, we see an opposite behavior: the volume at the `bounce' is maximum and the `bounce' turns out to be almost symmetric. Hence for very small masses, the bounce turns out to be a recollapse. This appears to be a universal feature of all the choices studied in this manuscript. Using the symmetric bounce condition it is straightforward to see that 
for the current choice of prescribing areas to the loops, there is no real allowed value of mass $M$. Thus, a symmetric bounce can be approached but never obtained in Corichi-Singh quantization \cite{cs-ks}. 

In the right plot in Fig.  \ref{fig:opt0} we show the value of the volume at the bounce as a function of the mass. As we see, for a fixed value of the mass, the volume is finite. However, there is a threshold mass where it tends to zero. This is so because for masses equal to or smaller than this threshold mass, there is actually no black hole. Therefore, if one trusts the effective description provided here for very small values of the mass, one finds the existence of a minimum nonzero mass in this scenario. It should be emphasized that the existence of a minimum non-zero mass of the black hole and a recollapse replacing the bounce for tiny black holes are novel results noticed here for the first time in loop quantization of Schwarzschild interior.

\subsection{Choice 1: $\beta=1$}\label{sec:case1}

From now on, we will study the symmetric bounce scenarios, starting with the choice $\beta=1$ in Eq. \eqref{eq:new-poly}. We obtain 
\begin{equation}
\delta_b =\alpha\frac{\sqrt{\Delta}}{r_o},\quad \delta_c =\frac{\sqrt{\Delta}}{L_o}.
\end{equation}

\begin{figure}
	\centering
	\includegraphics[width=0.49\textwidth]{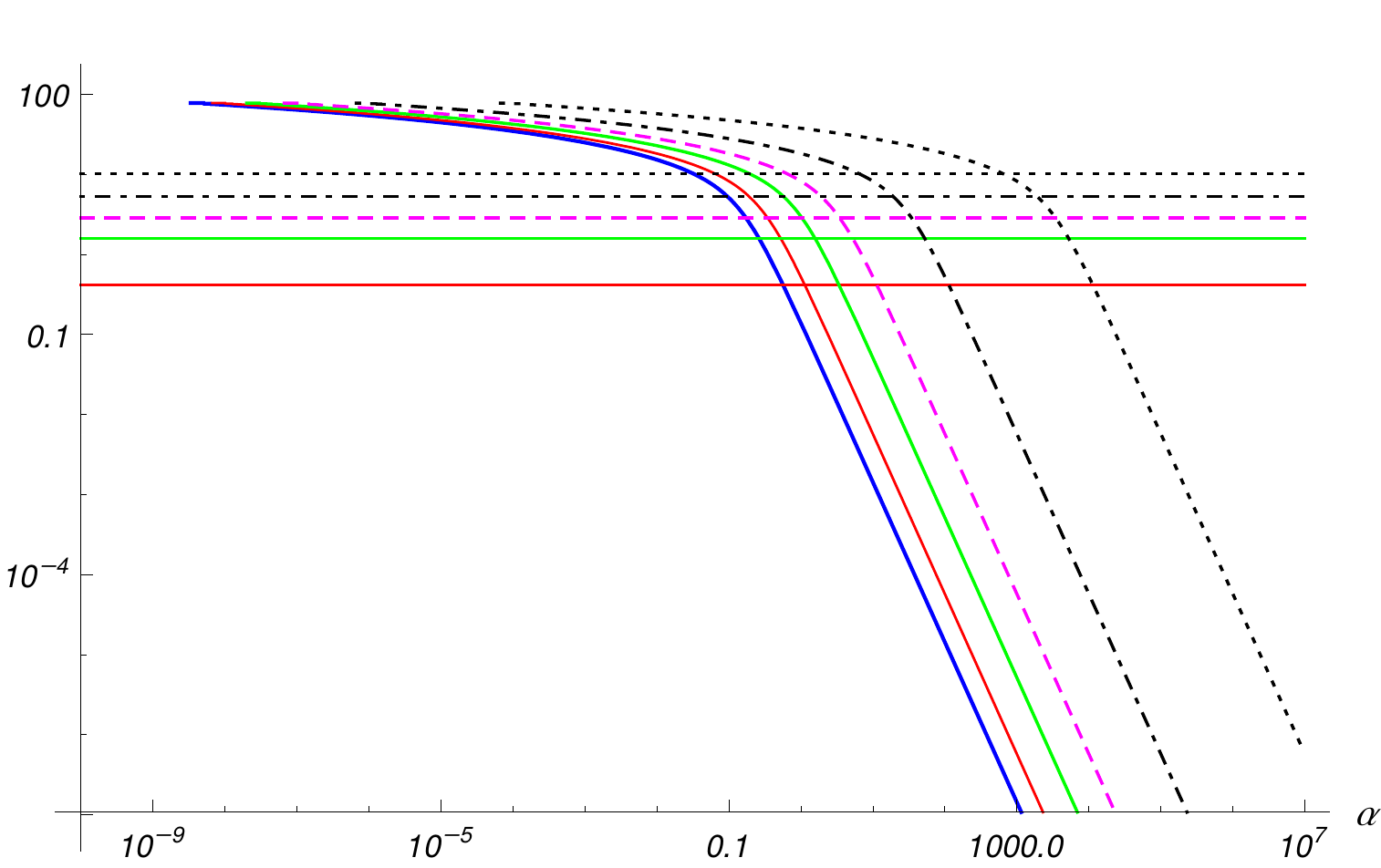} 
	\includegraphics[width=0.49\textwidth]{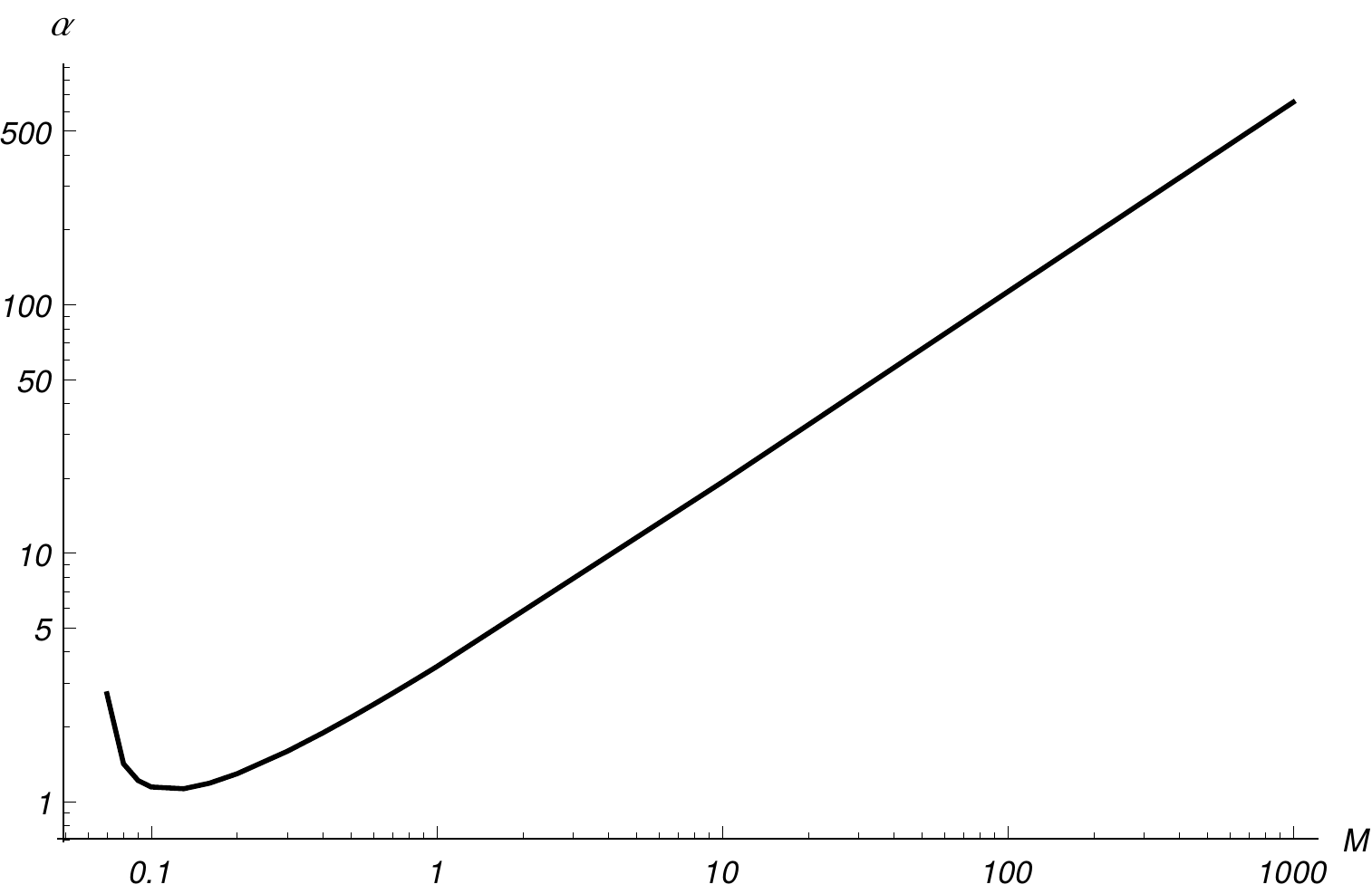} 
	\caption{In the left graph we plot the left hand side of Eq. \eqref{eq:opt1-alpha} as a function of $\alpha$ for different values of the mass $M$. The curves ranging from the thickest to the thinnest correspond to $M=0.1,0.3,1,10,1000$ in Planck units, respectively. The horizontal lines correspond to the right hand side of \eqref{eq:opt1-alpha}. The right plot provides $\alpha$ as a function of $M$ in the intersection points of the left graph.}
	\label{fig:opt1-alpha}
\end{figure}

If we substitute these polymer parameters in the symmetric bounce condition given by Eq. \eqref{eq:sym-bcond}, namely 
\begin{equation}\label{eq:opt1-alpha}
\frac{1}{G \gamma\left(\sqrt{1+\alpha^2  \frac{\gamma^2\Delta}{(2GM)^2}} \right)} {\rm arctanh}\left[\frac{1}{\sqrt{1+\alpha^2  \frac{\gamma^2\Delta}{(2GM)^2}}}\right]=\frac{1}{4 G \gamma}\log\left[\frac{8G M}{\gamma  \sqrt{\Delta}}\right],
\end{equation}
we can fix the parameter $\alpha$ in order to fulfill this condition. In order to find the allowed values of $\alpha$, in Fig. \ref{fig:opt1-alpha} we plot the left and right hand sides of Eq. \eqref{eq:opt1-alpha} separately, to find out the solutions given by their intersection points for different values of the mass. The right hand side, since it is independent of $\alpha$, is represented by the horizontal lines. As we see in the second graph of Fig. \ref{fig:opt1-alpha}, $\alpha$ has a minimum around $M=0.1$ (in Planck units), increasing on either side of this point. Let us also notice that for higher masses, $\alpha$ increases as a function of the mass following a power law.

\begin{figure}
	\centering
	\includegraphics[width=0.49\textwidth]{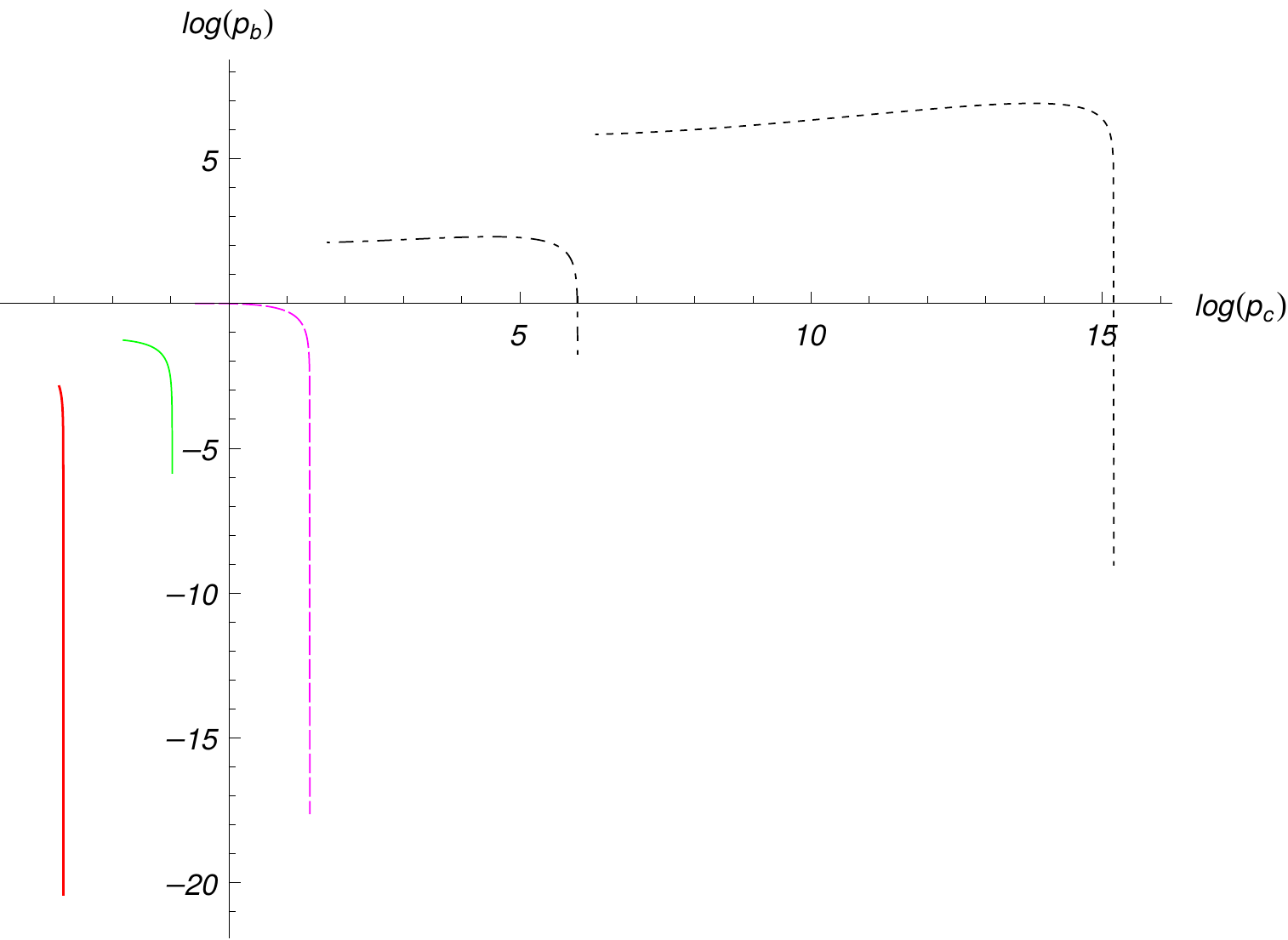} 
	\includegraphics[width=0.49\textwidth]{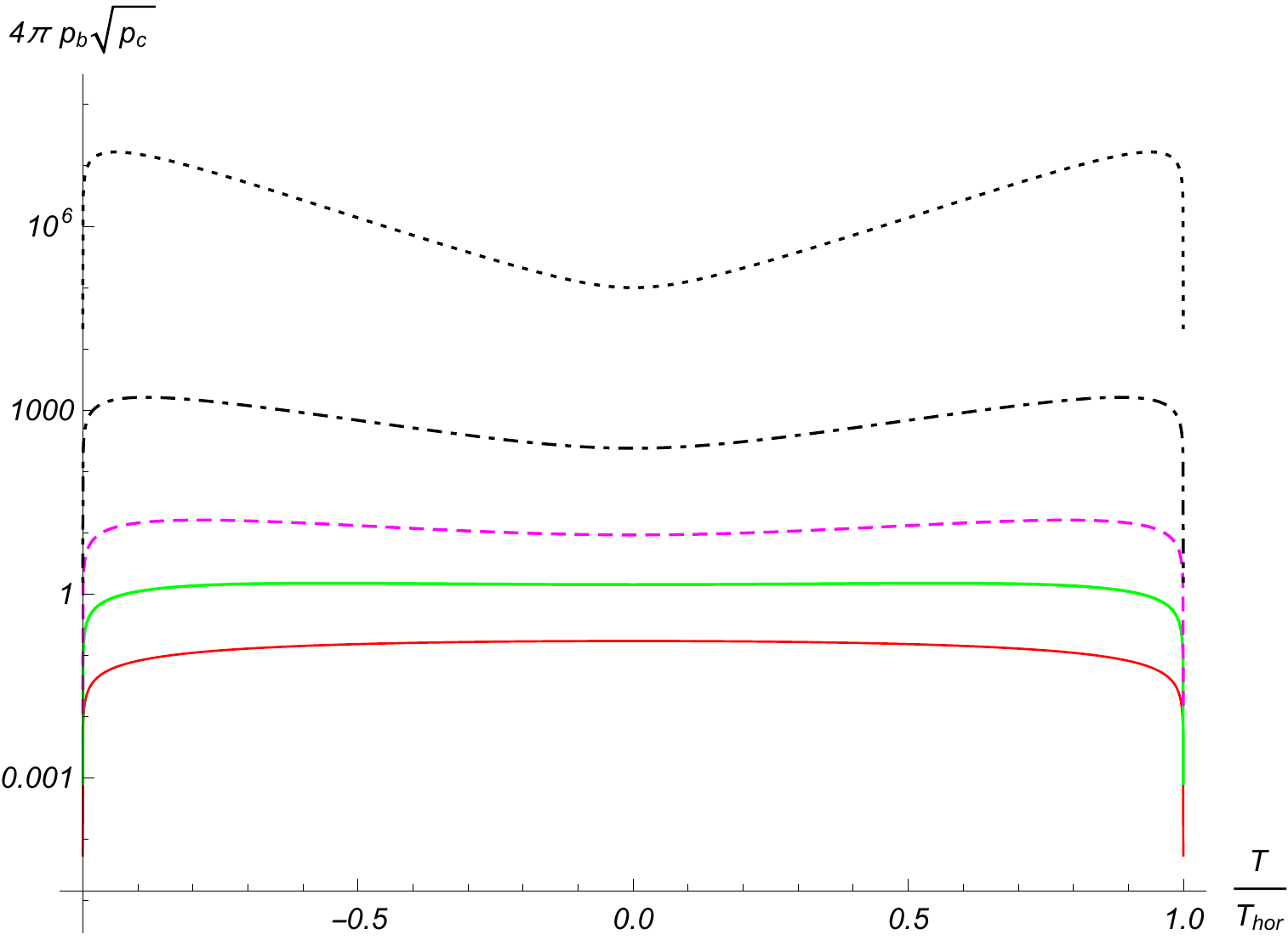} 
	\caption{In the left graph we plot the logarithmic of $p_b$ as a function of the logarithm of $p_c$ for different values of the mass $M$. The right graph provides the evolution of the spatial volume  as a function of the normalized time $T/T_{hor}$. The curves ranging from the thickest to the thinnest correspond to $M=0.05,0.1,0.3,1,10,1000$ in Planck units respectively.}
	\label{fig:opt1-triads}
\end{figure}

We have also plotted the dynamical evolution of the triads and the spatial volume in Fig. \ref{fig:opt1-triads}. As we see in the left graph, the value of $p_b$ goes to zero at the horizon. Regarding $p_c$, its value at the horizon is slightly higher than its classical value $(2GM)^2$. This is due to the small quantum corrections of the polymeric description. Besides, these curves enclose a vanishing area when ranging from horizon to horizon. This indicates that the bounce is symmetric. In the right graph we see that the spatial volume takes its absolute minimum value at the horizons (it vanishes there), as in the classical theory. For masses higher than $0.1$ it has a local minimum at the bounce, however those local minima become local maxima at the bounce for smaller masses. The bounce turns out to be a quantum recollapse for such small black hole masses. 

\begin{figure}
	\centering
	\includegraphics[width=0.49\textwidth]{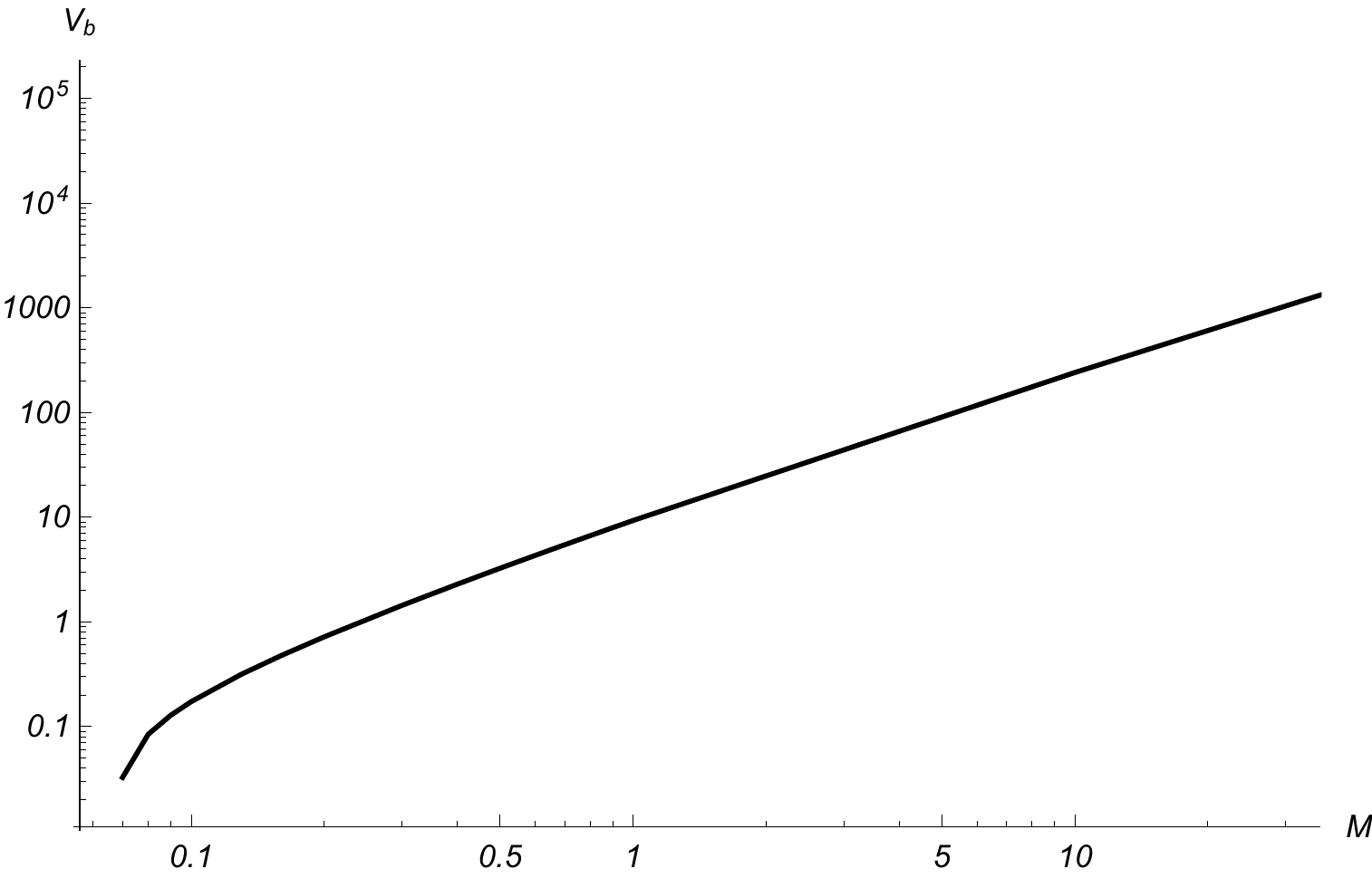} 
	\caption{Volume at the bounce as a function of mass for option $\beta=1$}
	\label{fig:opt1-bouncevolume}
\end{figure}

Besides, we plot the bounce volume versus the mass in Fig. \ref{fig:opt1-bouncevolume}. We see that it monotonically increases with the mass. We must notice that there is a change in the behavior around $M=0.1$. This corresponds to the turn around in the value of $\alpha$ as we go to lower masses, and for these masses the volume reaches local maxima at the bounce (instead of minima). The turn around in $\alpha$ as a function of $M$ does not lead to a turn around in the behavior of the volume a the bounce, but there is a change in the way the volume depends on the mass for these extremely low masses. We should note that for such a range of parameters we are essentially reaching the regime where effective spacetime description may not be well trusted, but in any case it will be interesting to gain more insights on this behavior. We find that this peculiar behavior for masses below $0.1$ is shared with all the other cases considered here.

\subsection{Choice 2: $\alpha = 1$}

We now proceed with another possible choice for the polymer parameters which allows symmetric bounce. We fix the length of the open loop as $\alpha=1$. Then, the polymer parameters take the form
\begin{equation}\label{eq:opt2-param}
\delta_b =\frac{\sqrt{\Delta}}{r_o},\quad \delta_c =\beta\frac{\sqrt{\Delta}}{L_o}.
\end{equation}

\begin{figure}
	\centering
	\includegraphics[width=0.49\textwidth]{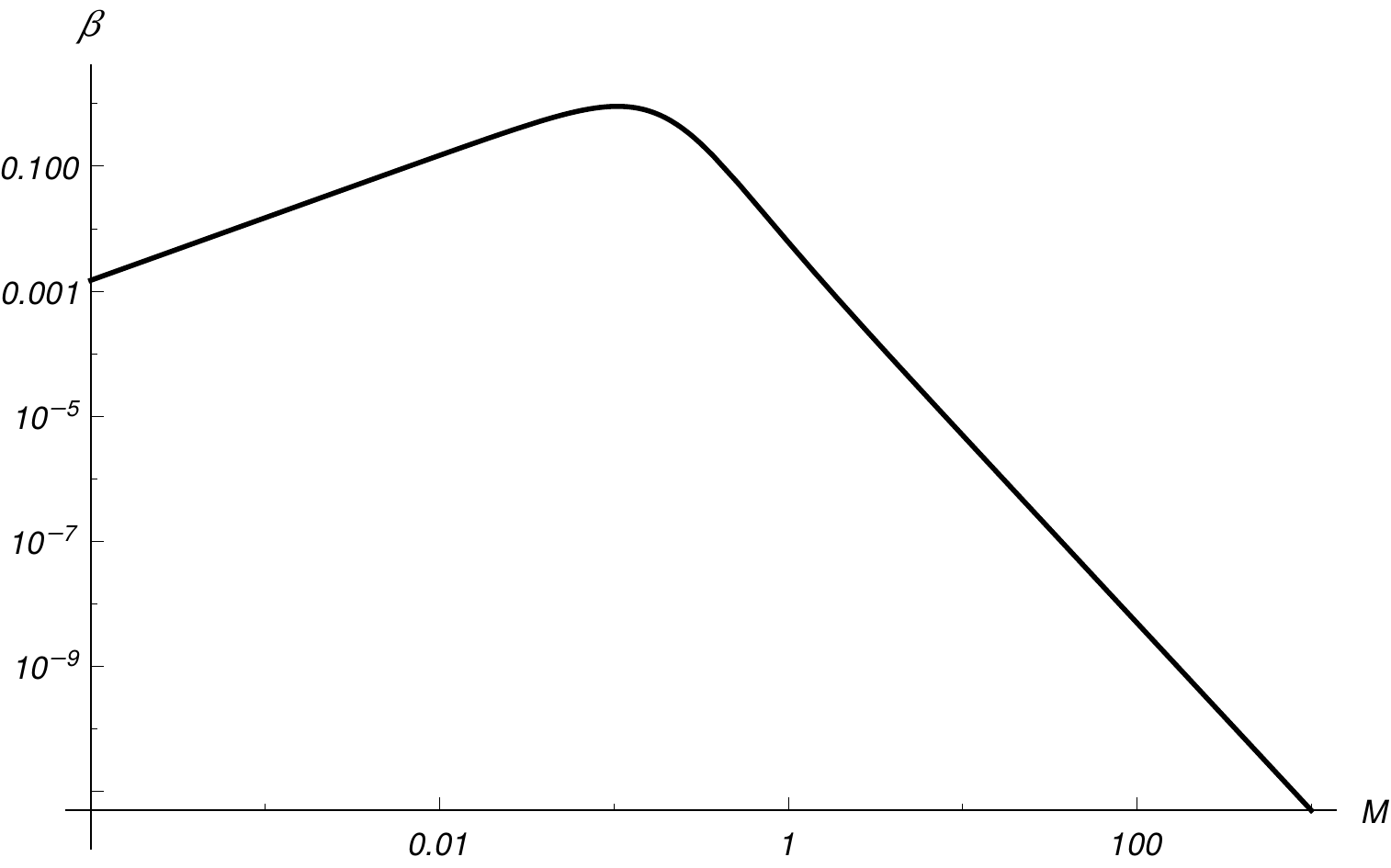} 
	\caption{The parameter $\beta$ is plotted as a function of the mass $M$ for choice 2.}
	\label{fig:opt2-betam}
\end{figure}

The loop with length determined by $\beta$, will be actually fixed by the symmetric bounce condition \eqref{eq:sym-bcond}
\begin{equation}\label{eq:opt2-beta}
\frac{1}{G \gamma\left(\sqrt{1+  \frac{\gamma^2\Delta}{(2GM)^2}} \right)} {\rm arctanh}\left[\frac{1}{\sqrt{1+ \frac{\gamma^2\Delta}{(2GM)^2}}}\right]=\frac{1}{4 G \gamma}\log\left[\frac{8G M}{\beta \gamma \sqrt{\Delta}}\right].
\end{equation}
Unlike the choice 1 for polymer parameters, this time this expression can be solved analytically for $\beta$, yielding
\begin{equation}\label{eq:opt2-betab}
\beta=\frac{8G M}{\gamma  \sqrt{\Delta}}\frac{(b_o-1)^{2/b_o}}{(b_o+1)^{2/b_o}},
\end{equation}
recalling that in this case 
\begin{equation}\label{eq:bo-opt2}
b_o=\sqrt{1+  \frac{\gamma^2\Delta}{(2GM)^2}}.
\end{equation}
It is useful to consider the  asymptotic behavior of $\beta$:
\begin{equation}\label{eq:beta-asym}
\lim_{M\to\infty} \beta=\frac{\gamma ^3 \Delta ^{3/2}}{32 G^3 M^3}+{\cal O}(M^{-5} \log[M]),\quad \lim_{M\to0} \beta=\frac{8 G M}{\gamma  \sqrt{\Delta }} +{\cal O}(M^{3}).
\end{equation}
In both cases (small and large black hole masses), the parameter $\beta$ goes to zero. In Fig. \ref{fig:opt2-betam} we plot $\beta$ as a function of the mass $M$. We see that it goes to zero for sufficiently small and large values of the mass, and its maximum is around $M=0{.}2$ (in Planck units).

\begin{figure}
	\centering
	\includegraphics[width=0.49\textwidth]{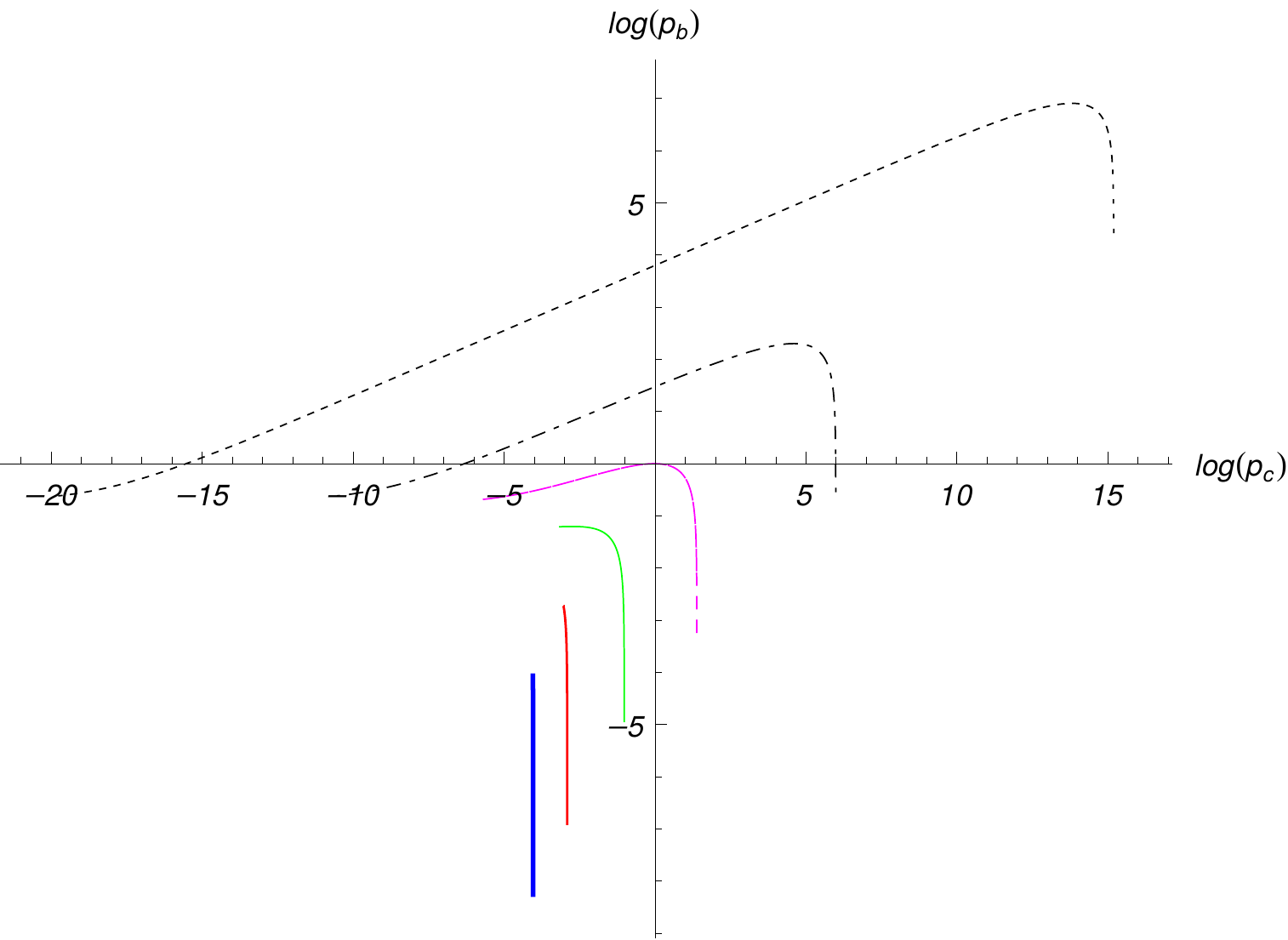} 
	\includegraphics[width=0.49\textwidth]{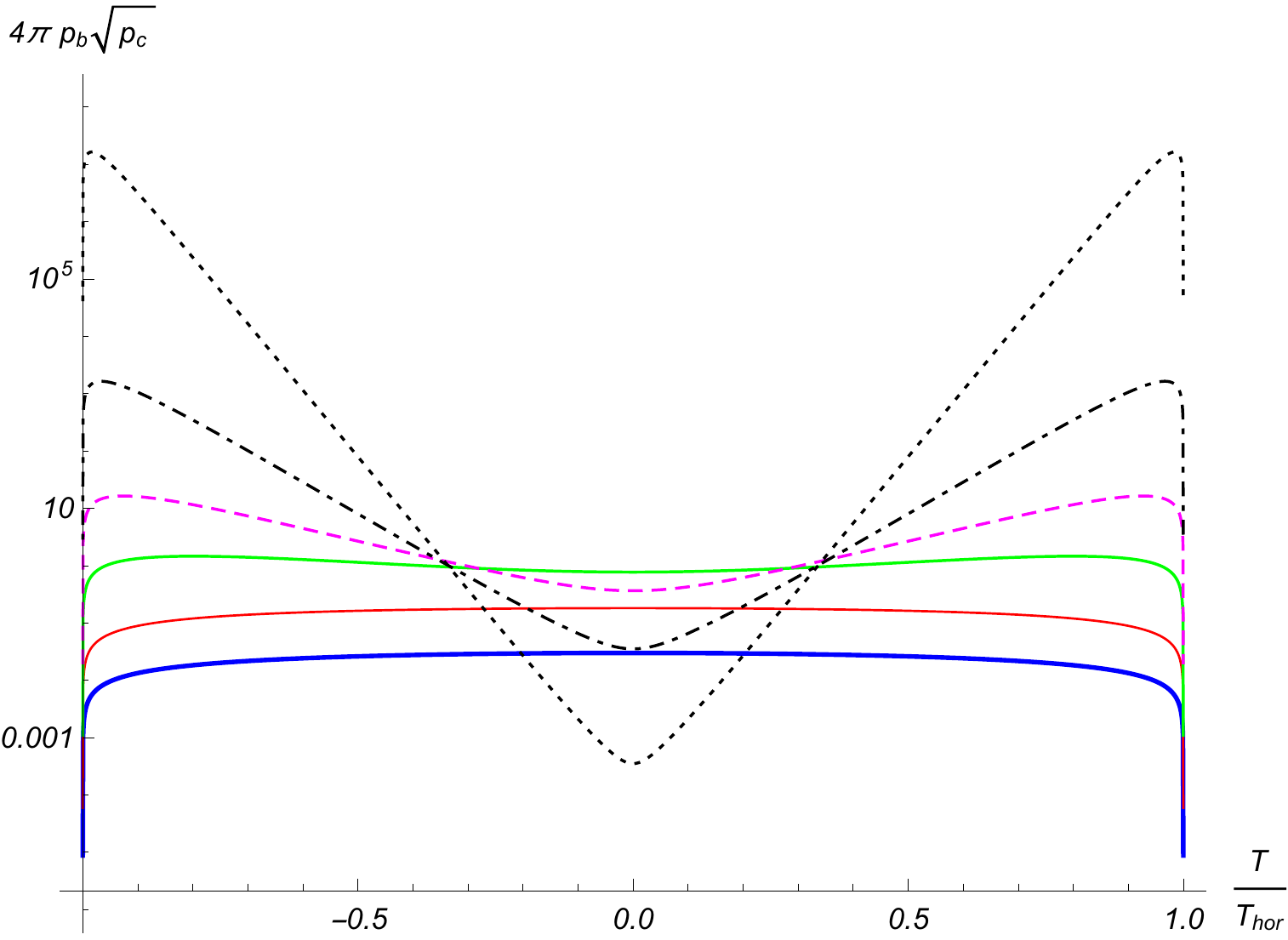} 
	\caption{In the left graph we plot the logarithmic of $p_b$ as a function of the logarithm of $p_c$ for different values of the mass $M$. The right graph provides the evolution of the spatial volume as a function of the normalized time $T/T_{hor}$. The curves ranging from the thickest to the thinnest correspond to $M=0.05,0.1,0.3,1,10,1000$, in Planck units, respectively.}
	\label{fig:opt2-triads}
\end{figure}

\begin{figure}
	\centering
	\includegraphics[width=0.49\textwidth]{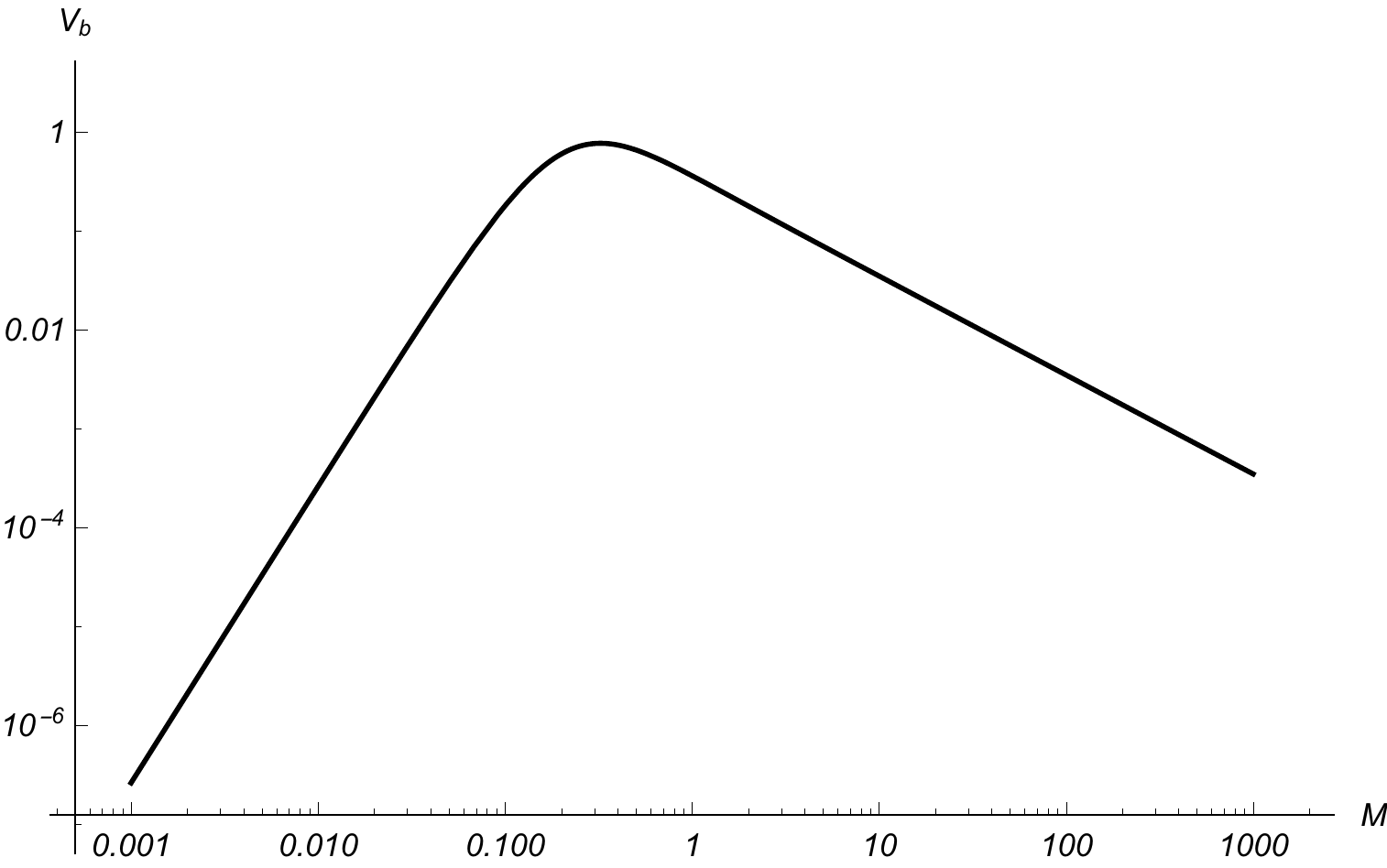} 
	\caption{Bounce volume as a function of mass for case $\alpha=1$.}
	\label{fig:opt2-bouncevolume}
\end{figure}

Regarding the triad components, analytical expressions can be provided as functions of $T$ and the mass $M$. We do not give here the expressions, but one can obtain them by replacing in Eqs. \eqref{eq:cb-to-T-lqc} the Eqs. \eqref{eq:opt2-param}, \eqref{eq:opt2-betab} and \eqref{eq:bo-opt2}. In Fig. \ref{fig:opt2-triads} we plot the dynamical evolution of the triads and the spatial volume. As we see in the left graph $p_b$ goes to zero at the horizon and $p_c$ is again slightly higher than its classical value $(2GM)^2$, as it is expected from the contribution of the small quantum corrections. The triads $p_c$ and $p_b$ evaluated at the bounce take the form
\begin{equation}
p_c(T=0)=\gamma\sqrt{\Delta} GM\beta(M),\quad p_b(T=0)=\frac{L_o\gamma  \sqrt{\Delta }}{1+\frac{\gamma ^2 \Delta }{4 G^2M^2}}.
\end{equation}
Then, we see that, from Eq. \eqref{eq:beta-asym}, at large values of the mass $M$, the triad $p_c$ at the bounce decreases quadratically (at leading order) as a function of the mass while $p_b$ converges to the constant value $L_o\gamma  \sqrt{\Delta }$. On the other hand, for very small mass $M$, both $p_c$ and $p_b$ at the bounce behave, at leading order, as quadratic functions of the mass. In the right graph we show the behavior of the spatial volume. It takes its minimum absolute value at the horizons, since it vanishes there. Just like the previous cases, the volume reaches a local minimum at the bounce if the mass is higher than $0.1$,  but for smaller values of the mass the volume reaches a local maxima. However, unlike the previous case, the turn around in the value of $\beta$ causes a turn around in the bounce volume as well. This is shown in Fig. \ref{fig:opt2-bouncevolume} where we show the volume at the bounce versus the mass $M$. We see that it is no longer a monotonic function of the mass.

\subsection{Choice 3: $\beta=1/\alpha$}

\begin{figure}
	\centering
	\includegraphics[width=0.49\textwidth]{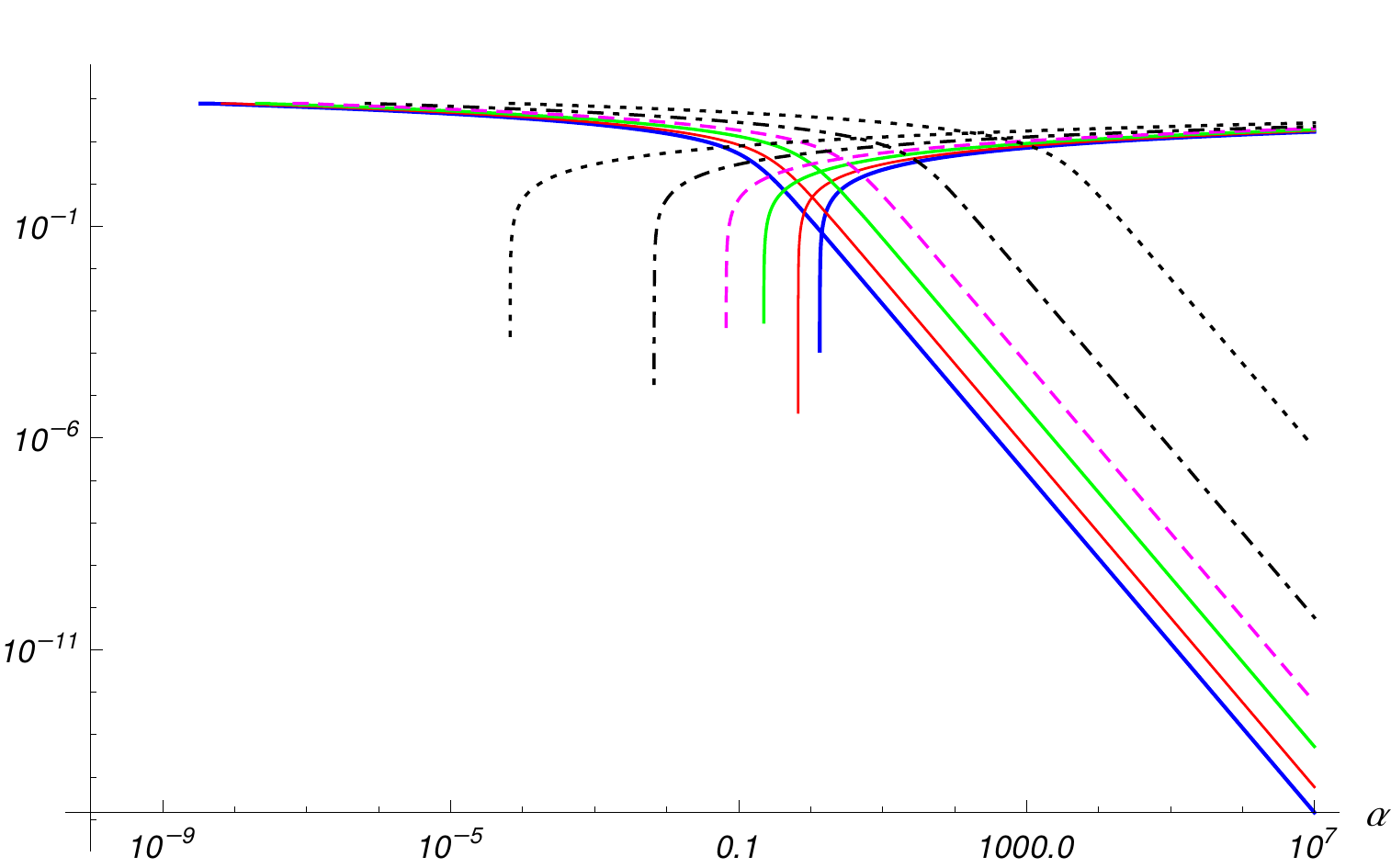} 
	\includegraphics[width=0.49\textwidth]{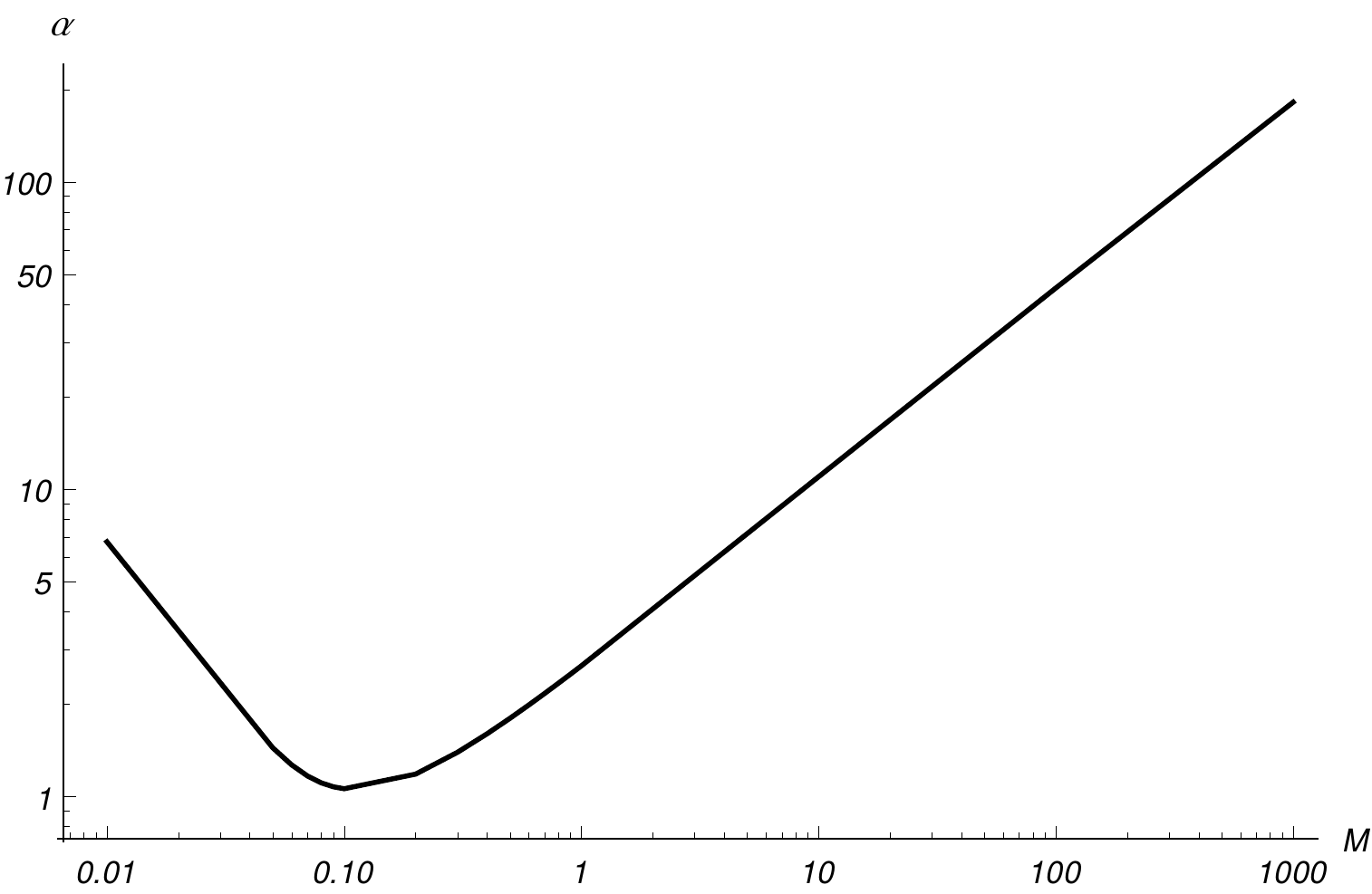} 
	\caption{In the left graph we plot the left and right hand side of Eq. \eqref{eq:opt3-1/alpha} as a function of $\alpha$ for several values of the mass $M$. The curves ranging from thickest to the thinnest correspond to $M=0.05,0.1,0.3,1,10,1000$ (in Planck units), respectively. The curves converging on the left of the graph are of the left hand side of Eq. \eqref{eq:opt3-1/alpha}, and vice-versa. The right graph provides $\alpha$ as a function of $M$ in the intersection points of the left graph.}
	\label{fig:opt3-alpha1}
\end{figure}

Now we will proceed with the most interesting choice among the ones considered here. The reason is that this choice allows us to work with a closed loop since, from Eq. \eqref{eq:new-poly} we get
\begin{equation}
\delta_b r_o \delta_c L_o=\Delta,
\end{equation}
in partial agreement with Ref. \cite{cs-ks}. The choice  $\beta=1/\alpha$ amounts to
\begin{equation}
\delta_b =\alpha\frac{\sqrt{\Delta}}{r_o},\quad \delta_c =\frac{\sqrt{\Delta}}{L_o \alpha}.
\end{equation}
Together with the symmetric bounce condition \eqref{eq:sym-bcond}
\begin{equation}\label{eq:opt3-1/alpha}
\frac{1}{\sqrt{1+\alpha^2  \frac{\gamma^2\Delta}{(2GM)^2}}} {\rm arctanh}\left[\frac{1}{\sqrt{1+\alpha^2  \frac{\gamma^2\Delta}{(2GM)^2}} }\right]=\frac{1}{4}\log\left[\frac{8G M \alpha}{\gamma  \sqrt{\Delta}}\right],
\end{equation}
we can fix $\alpha$ as in the previous sections. Unlike the case of choice 2, we do not have a closed form expression for  $\alpha$. To find the allowed values of $\alpha$, in Fig. \ref{fig:opt3-alpha1} we plot the left and right hand sides of Eq. \eqref{eq:opt3-1/alpha} separately for different values of the mass $M$. The intersection points show that $\alpha$ grows monotonically with the mass of the black hole $M$, just like for the case $\beta=1$. 

\begin{figure}
	\centering
	\includegraphics[width=0.49\textwidth]{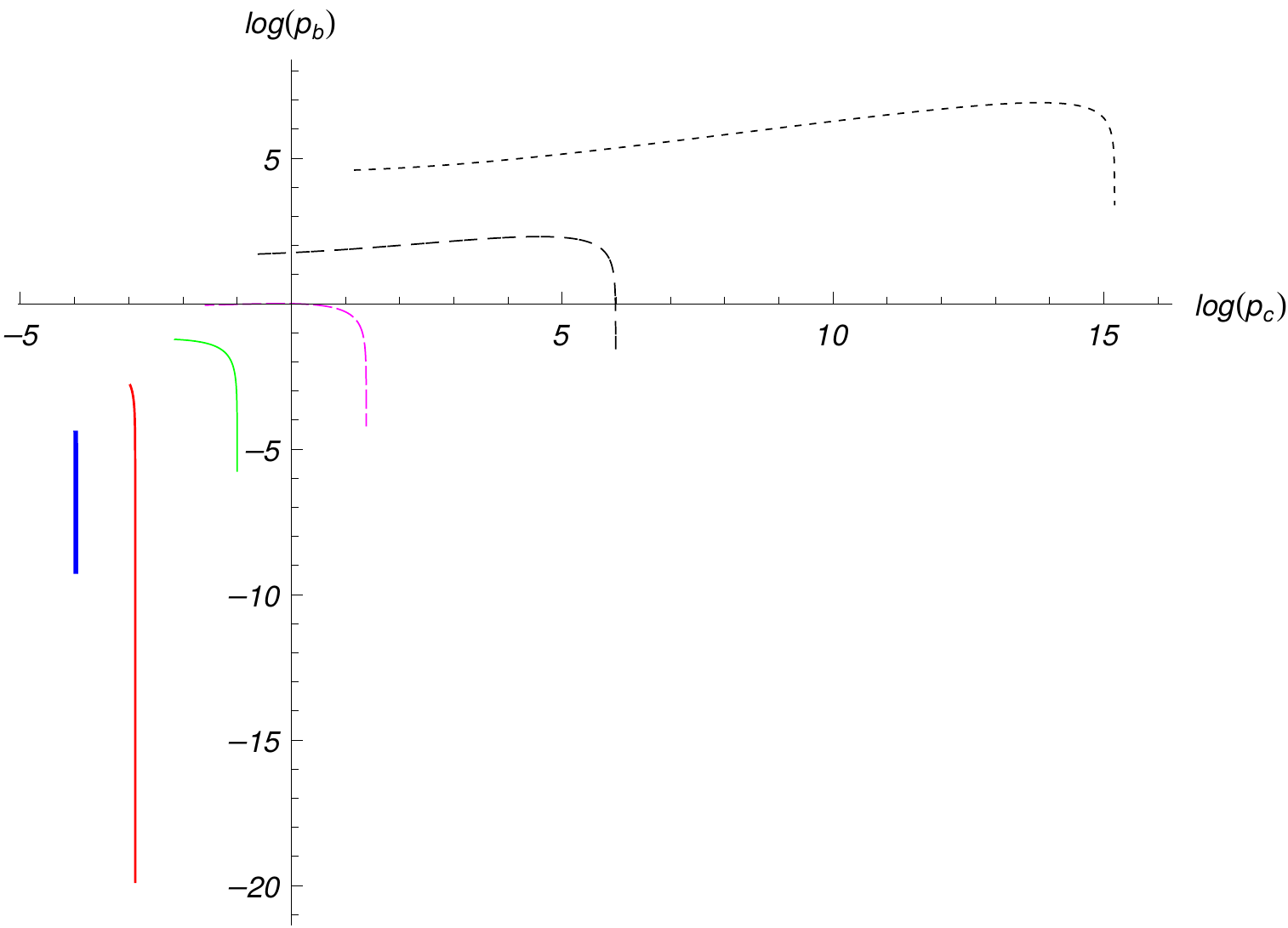} 
	\includegraphics[width=0.49\textwidth]{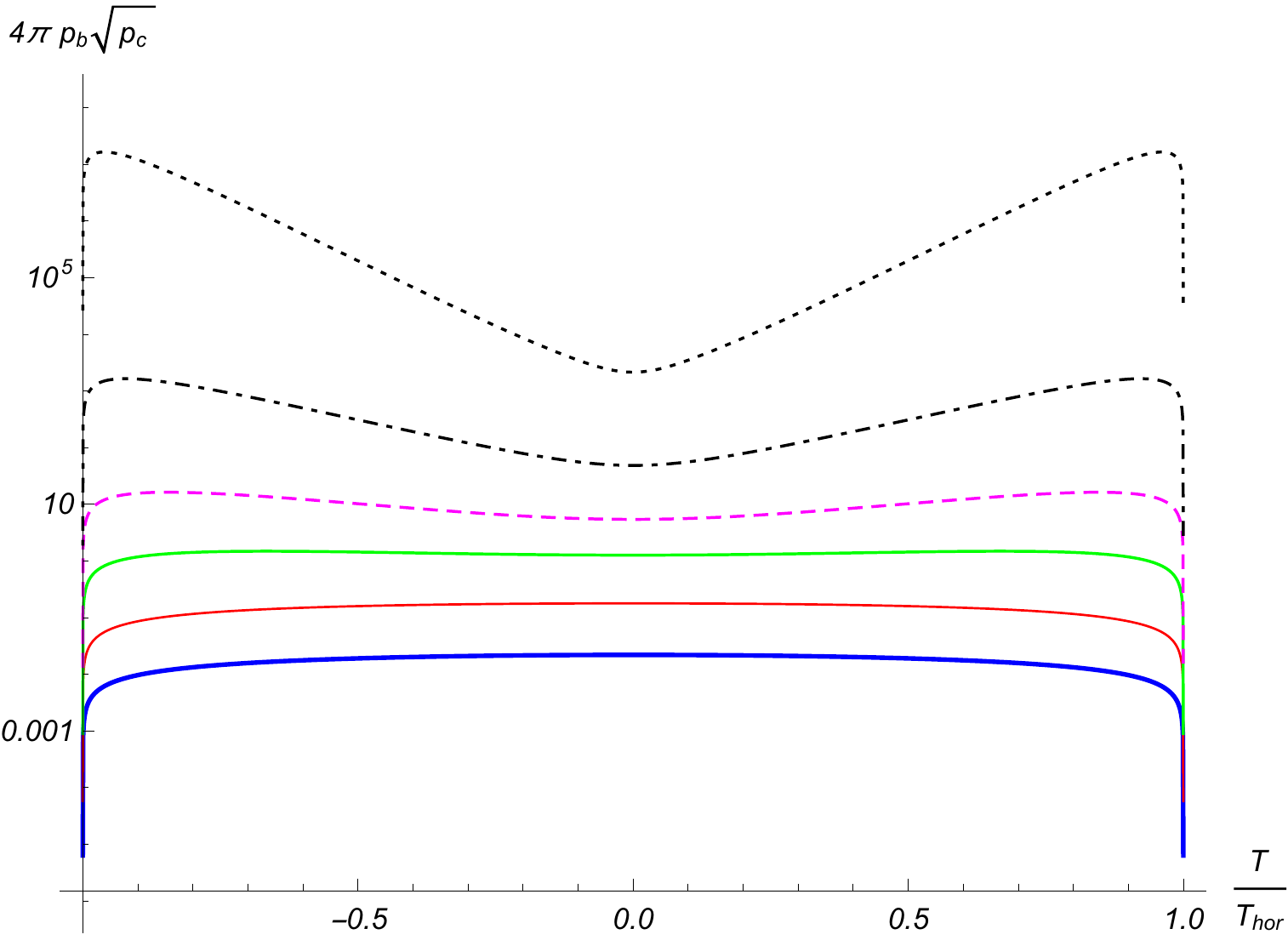} 
	\caption{In the left graph we plot the logarithmic of $p_b$ as a function of the logarithm of $p_c$ for different values of the mass $M$. The right graph provides the evolution of the spatial volume as a function of the normalized time $T/T_{hor}$. The curves ranging from the thickest to the thinnest correspond to $M=0.05,0.1,0.3,1,10,1000$, in Planck units, respectively.}
	\label{fig:opt3-triads}
\end{figure}

\begin{figure}
	\centering
	\includegraphics[width=0.49\textwidth]{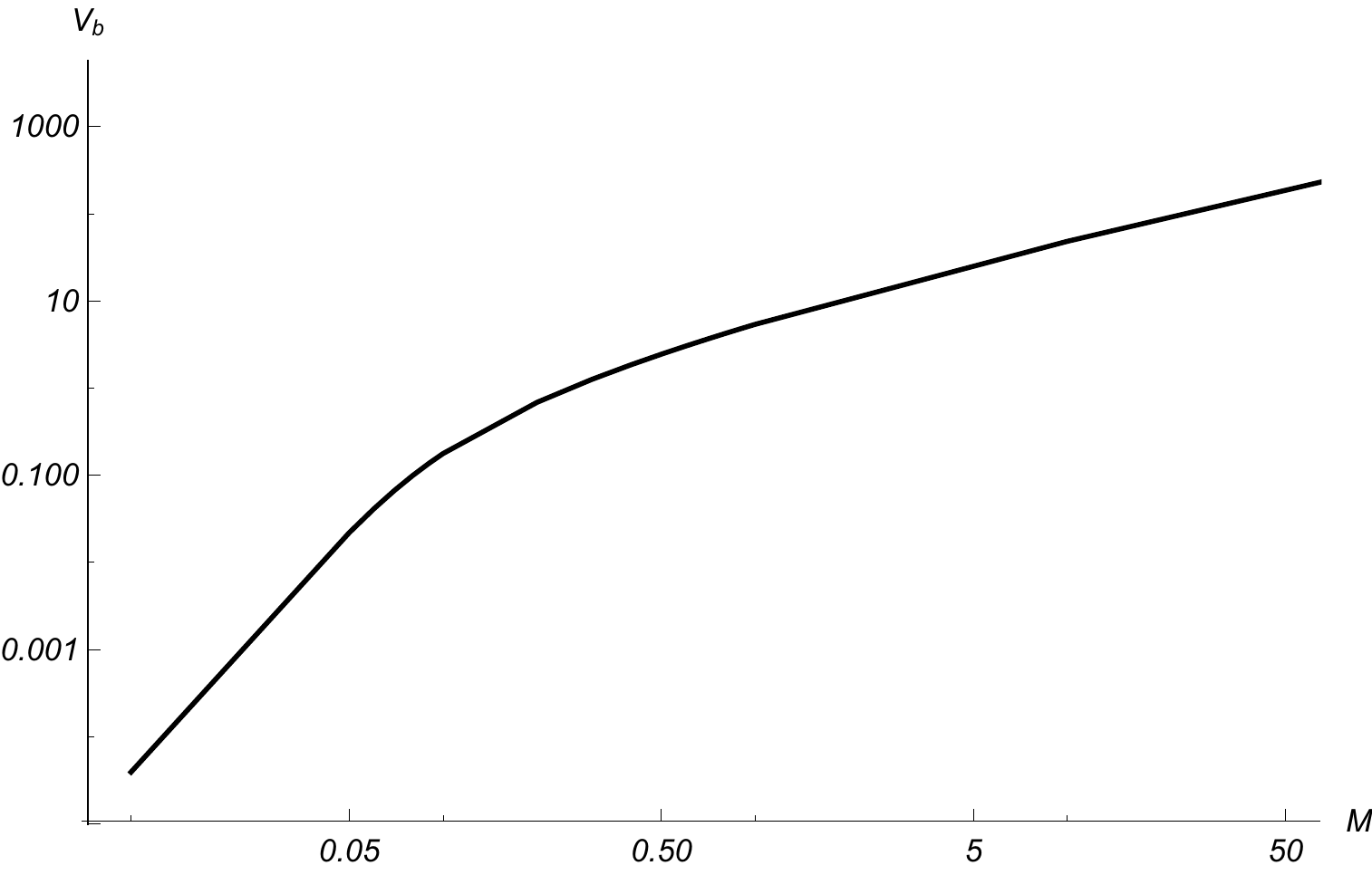} 
	\caption{Bounce volume as a function of mass for case $\alpha=1$.}
	\label{fig:opt3-bouncevolume}
\end{figure}

We note the striking similarity between the graphs obtained for the variation in $\alpha$ with $M$ in the present case with the case $\beta=1$. First note that the left hand side of the symmetric bounce condition is identical in both cases and hence generate identical curves on the graph. In the present case, even though the plots for the right hand side look quite different from the case $\beta=1$ (choice 1), the behavior of $\alpha$ as a function of $M$ has similarities with respect to the case $\beta=1$, as shown in the left graph of Fig. \ref{fig:opt3-alpha1}. We also find that the behavior of the triads and the spatial volume in the present case are similar to the case $\beta=1$, as shown in Fig. \ref{fig:opt3-triads}. There, we have plotted the dynamical evolution of the triads and the spatial volume. As we see in the left graph, the value of $p_b$ goes to zero at the horizon. Regarding $p_c$, due to quantum geometry corrections, its value at the horizon is slightly higher than its classical value $(2GM)^2$, as in the previous cases. The behavior of the spatial volume as a function of time and mass is very similar to the situation in choice 1. It shows the same peculiar behavior where the spatial volume for masses lower than a certain value is a convex function of time and reaches a global maximum (instead of a local minimum) at the bounce. This transition from normal to abnormal behavior is clearly indicated by the kink in the graph of bounce volume versus mass shown in Fig. \ref{fig:opt3-bouncevolume}, where we show the value of the spatial volume at the bounce as a function of the mass $M$. It turns out to be a monotonically growing function of the mass $M$. 

\begin{figure}[h]
	\centering
	\includegraphics[width=0.49\textwidth]{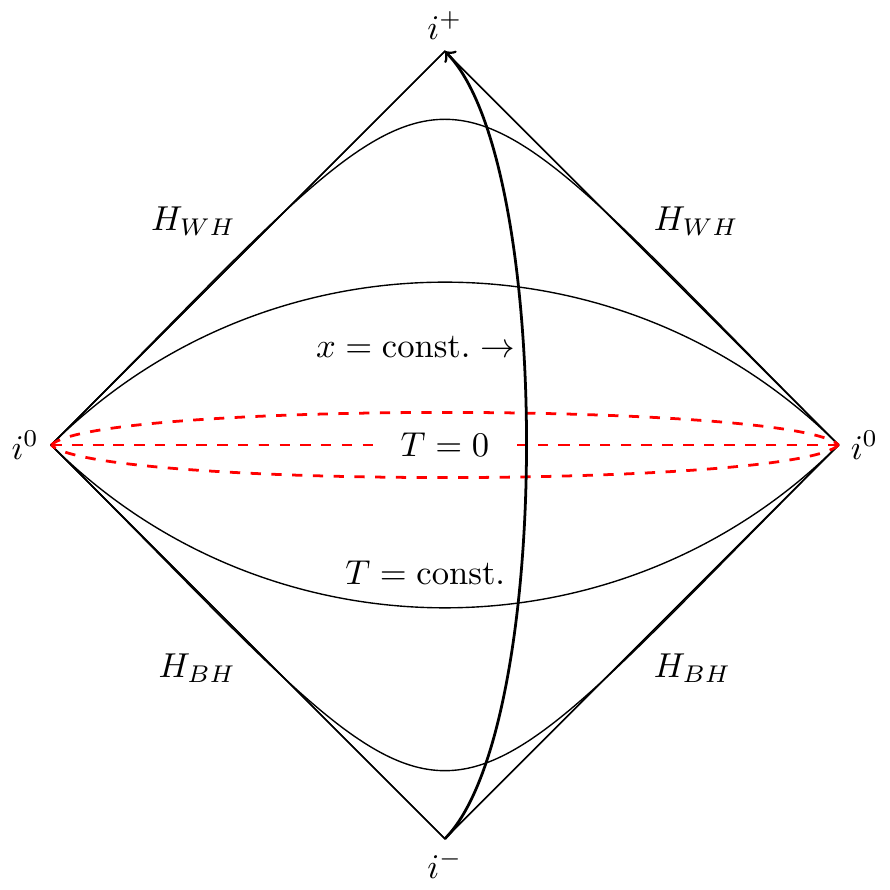} 
	\caption{Penrose diagram for the interior of Schwarzschild spacetime undergoing symmetric bounce to white hole solution is shown.}
	\label{fig:p-diag}
\end{figure}

Finally, for completeness, in Fig. \ref{fig:p-diag} we include a Penrose diagram for these spacetimes.
We only show the interior spacetime as the quantizations studied in this manuscript do not yet include the exterior spacetime.  In the Penrose diagram, the black and white hole horizons are denoted by $H_{BH}$ and $H_{WH}$, respectively. Horizontal curves represent Cauchy surfaces ($T={\rm const.}$) while vertical curves denote $x={\rm const.}$ observers. As usual, $i^\pm$ represent $T\to\pm T_{\rm hor}$, while $i^0$ denotes $x=0$ and $x=L_0$ (notice the symmetry of the diagram). A test observer which falls from the black hole horizon passes through the
quantum gravitational regime, represented by red (shaded) dashed curves, and reaches the white hole horizon without encountering the classical singularity. Due to quantum geometric effects, the latter is removed from the
effective spacetime. Due to the symmetric bounces in prescriptions I, II and III, the $T=0$ curve which
represents the bounce time lies in the middle of the dashed regime. For the prescription in Corichi-Singh model, this region is not expected to be symmetric. 

\section{Conclusions}
Unlike the classical theory where the evolution in cosmological and gravitational collapse scenarios ends in a singularity, there is a growing evidence from various investigations over the last decade, that modifications 
from loop quantum gravity robustly lead to resolution of singularities. Quantum bounce has been a generic feature of all the spacetimes 
studied so far, starting from the first results in isotropic and homogeneous settings in LQC \cite{aps2,aps3,acs,mmo,mop} to the black hole interior as 
discussed in this manuscript. However, depending on the spacetime the structure of the singularity and of the bounce can be quite 
different especially when the spacetime has non-zero anisotropies. The non-vanishing Weyl curvature results in an anisotropic approach to singularity 
causing the bounce to be asymmetric. In the Schwarzschild interior, which is isometric to the Kantowski-Sachs vacuum spacetime, the 
situation has so far been believed to be similar. In the only known quantization so far in which physics does not depend on the fiducial cell and which 
yields a well defined infra-red limit as GR, the bounce from the black hole to the white hole turns out to be highly asymmetric. So much 
that the mass of the child white hole is a quartic power of the mass of the mother black hole \cite{cs-ks}. The main objective of our analysis was to understand 
whether a symmetric bounce is possible in this quantization or its modification without spoiling any of its nice features. 

To answer the above question we used the Dirac's method for constrained theories. After identifying (partial) Dirac observables we wrote the Hamiltonian constraint in terms of them and performed a gauge fixing to identify a geometrical clock. This approach allows us to extract a symmetric bounce condition in terms of the Dirac observables and the internal clock. One of the advantages of this strategy is that the model can be eventually quantized within a reduced phase space quantization. Though we do not perform a quantization, we study the physical implications within the classical and the effective loop dynamics. Our analysis can be therefore viewed as a first concrete step towards 
exploring the viability of a quantization yielding a symmetric bounce. 

The main result of our analysis is that symmetric bounce is indeed possible  by slightly modifying the quantization prescription put forward in Ref. \cite{cs-ks}. 
If the assignment of the minimum area of the loops over which holonomies are considered has some freedom, which we parameterized using $\alpha$ and $\beta$ in Sec. IV, then given the 
mass of the black hole and using effective dynamics one can choose $\alpha$ and $\beta$ such that the child white hole mass turns out to be identical to the mother black hole mass. 
Among the three choices considered which yield symmetric bounce, choice 3 is closest to the Corichi-Singh quantization and the most frugal choice because it only exploits a 
remaining freedom in the original quantization which deals with the assignment of area to the open loop. However it is important to note that the freedom considered in all the choices may have origins 
in the full theory. In loop quantum gravity, using coherent state techniques there is already evidence \cite{liegener} that the cosmological sector corresponds not exactly to LQC but a modification which 
can be understood as arising from difference in assigning the area to the loops over which holonomies are considered. If such a result holds for the Kantowski-Sachs vacuum spacetime then all the considered 
choices would be potentially viable. 

We find that the symmetric bounce condition can never be fulfilled in the Corichi-Singh quantization for any real value of the black hole mass. 
For the small masses the bounce tends to be symmetric but an exact symmetry can not be achieved. It is notable that the features of the effective dynamics for very small masses (less than Planck mass) are essentially similar for all the choices considered in this manuscript. In all of the choices the bounce volume is a non-linear function of the black hole mass, and a non-monotonic function in choice 2 considered in Sec. IV.
We find some novel results from the effective dynamics for all the choices including the Corichi-Singh quantization. On the one hand, there exists a smallest mass, approximately $M \sim 0.1$, below which effective dynamics does not permit a black hole solution. This is actually in good agreement with recent quantizations of the full spherically symmetric spacetime \cite{gp-BH-1,gp-BH-2}. On the other hand, for small masses such that the Schwarzschild radius is of the order or smaller than the discreteness scale, the bounce in the interior is replaced by a recollapse. 
This is a purely quantum behavior which has no classical analog for this particular spacetime. It might happen that the effective dynamics studied here is not valid anymore. In any case, this is an intriguing result which needs to be investigated further in a future work. Apart from these generic results, we find non-trivial relationships between the behavior of the parameters $\alpha$ and $\beta$ with the mass of the black hole for different choices of the area of the loop assignment. The behavior of the volume at the bounce also shows a non-trivial behavior as function of the mass of the black hole. These relationships also need to be carefully understood to gain more insights on each of the choices. It is to be emphasized that all these results are new and quite unanticipated from the previous studies of loop quantization of the Kantowski-Sachs spacetime.

It is important to comment on one of the main technical differences of our analysis which distinguishes it from the Corichi-Singh quantization \cite{cs-ks}. As in the Corichi-Singh quantization the areas of the loops depend on the mass of the black hole the important difference is that in our analysis one needs to know about the effective dynamics in order to find suitable values of the quantization parameters $\alpha$ and $\beta$ to find a symmetric bounce. Unlike the Corichi-Singh quantization where, given the mass of a black hole a quantum theory can be constructed, in our treatment the quantization parameters need to be tuned to a particular value -- an exercise which depends on solving the effective Hamiltonian constraint. This requires a two step process. In the first step, given a black hole mass in a generalized Corichi-Singh quantization with to-be-fixed quantization ambiguity parameters, we quantize the model and then deduce the effective dynamics. We then solve for the values of $\alpha$ and $\beta$ requiring the symmetric bounce condition. We eventually repeat the quantization procedure with these particular choices of $\alpha$ and $\beta$. This procedure yields the desired loop quantization yielding equal masses of mother black hole and child white hole. In a way the quantization choices we presented here result in an improvement procedure over Corichi-Singh quantization to find a symmetric bounce. Let us also mention that our results are a proof of existence of symmetric bounces in these scenarios rather than a consequence derived from first principles. It would be interesting to study in the future if these choices can actually be obtained from deeper quantum geometry effects.

Existence of the non-singular symmetric quantum gravitational bounce in the Schwarzschild interior as found in this manuscript provides a root to various phenomenological ideas on black hole to white hole transition \cite{barcelo0,barcelo1,barcelo2,planckstar1,planckstar2} which generally assume a symmetric quantum gravity evolution. We have explicitly demonstrated that the quantum gravitational regime near the singularity can result in a symmetric bounce even if the singularity is dominated by the Weyl curvature and is highly anisotropic. Nevertheless, let us mention that this picture is not fully compatible with those of Refs.  \cite{barcelo0,barcelo1,barcelo2,planckstar1,planckstar2}  since for macroscopic black holes quantum gravity corrections seem to be confined within the high curvature regimes (close to the horizon we recover GR in a very good approximation). Still our work provides a platform for these studies and the understanding of their phenomenological aspects more rigorously \cite{wh-phen}. In fact using the techniques presented in our analysis it is quite straightforward to offset the symmetry of the bounce using quantization parameters $\alpha$ and $\beta$. Furthermore, the technique can be replicated in various different spacetimes which are loop quantized. As an example, it will be interesting to see whether the asymmetric bounces in Bianchi spacetimes can be made symmetric using similar prescriptions. Finally, our work provides a rigorous stage for phenomenological explorations in quantum gravitational black hole physics where symmetric bounces have played a central role.  It remains to be seen how such phenomenological studies can be linked to future observations.  

%Therefore, our proposal provides a new avenue
%to rigorously study quantum gravity effects in black hole physics with special interest in phenomenological explorations. 

\begin{acknowledgments}
This work is partially supported by NSF grants PHY-1404240 and PHY-1454832. J. O. acknowledges the partial support by NSF grants PHY-1305000 and PHY-1505411, the Eberly research funds of Penn State University (USA), the grant MICINN FIS2014-54800-C2-2-P (Spain), and Pedeciba (Uruguay).
\end{acknowledgments}

\appendix

\section{Gauge fixing $p_1=\tilde \tau$ in classical general relativity}\label{app:A}

In this Appendix we provide an alternative gauge fixing, with respect to the one adopted in Sec. \ref{sec:class-dyn}, corresponding to the second class condition $\tilde\Phi=p_1-\tilde \tau\approx 0$, where $\tilde \tau$ plays now the role of time, such that it takes any value in the negative real line (since $p_1\leq 0$). Preservation of this condition upon evolution 
\begin{align}
0\approx\dot{\tilde{\Phi}}=\{p_1,H_{\rm class}\}-1,
\end{align}
determines, on shell, the lapse function
\begin{equation}
	N=\frac{\gamma}{8 \pi \sqrt{1 - e^{2 G \gamma \tau}}} e^{G \gamma (3\tau+p_2^0)} \sqrt{\frac{|o_2|}{2}}.
\end{equation}
Besides, we have $o_1\approx-o_2$. The reduced action is now given by
\begin{equation}\label{eq:act-P2}
	S_{\rm red} = \int d\tilde \tau \left(p_2\dot{o}_2-\tilde h_{\rm red}\right),
\end{equation}
with $\tilde h_{\rm red}=-o_2$. The dynamics can be easily solved as in the gauge fixing adopted in Sec. \ref{sec:class-dyn}. It turns out that $o_2$ is a constant of the motion and $p_2=\tilde\tau+p_2^0$, with $p_2^0$ another constant of the motion. 

As before, we can express the observables $(o_2,p_2^0)$ in terms of the mass of the black hole and the fiducial length through the conditions at the horizon $p_b(\tilde \tau_{\rm hor})=0=b(\tilde \tau_{\rm hor})$, $p_c(\tilde \tau_{\rm hor})=(2 G M)^2$ and $c(\tilde \tau_{\rm hor})=\frac{\gamma L_o}{4 G M}$. 

Therefore, using again Eqs. \eqref{eq:c-to-Oi} and \eqref{eq:b-to-Oi}, one can see that $\tilde \tau_{\rm hor}=0$, and $o_2$ and $p_2^0$ as 
\begin{equation}
	p_2^0=\frac{1}{4\gamma G}\log\left(\frac{4GM}{\gamma L_o}\right),\quad o_2= 2GM\gamma L_o.
\end{equation}
Hence, $o_2$ and $p_2^0$ are completely determined by the mass $M$ of the black hole. Besides, in this gauge fixing, triads and connections as functions of time take the form
\begin{align}
	c(\tilde \tau) &=  \frac{\gamma L_o}{4GM}e^{-4 G \gamma  \tilde \tau},\quad p_c(\tilde \tau)=(2GM)^2e^{4 G \gamma \tilde \tau}, \\
	b(\tilde \tau) &= - \gamma e^{-G \gamma \tilde \tau} \sqrt{1 - e^{2 G \gamma \tilde \tau}}, \quad p_b(\tilde \tau) = 2GML_oe^{G \gamma \tilde \tau} \sqrt{1 - e^{2 G \gamma \tilde \tau}}.
\end{align}
Here, we consider again the sector where $p_b(\tilde \tau)\geq 0$ for all $\tilde \tau$, which involves $\epsilon_c=1$ and $\epsilon_b=-1$. Besides, the time $\tilde \tau\in(-\infty,0]$ and the singularity is located at $\tilde \tau\to-\infty$.

\section{Gauge fixing condition $P_1=\tilde T$ in effective loop quantum cosmology}\label{app:B}

Another gauge fixing condition that can be adopted within the effective dynamics in loop quantum cosmology corresponds to the second class condition $\tilde \Psi = P_1-\tilde T\approx 0$. Preservation of this condition upon evolution 
\begin{align}
0\approx \dot{\tilde \Psi} = \{P_1,H_{\rm LQC}\}-1,
\end{align}
and after evaluation on shell, allows us to fix the lapse function
\begin{equation}
N=\epsilon_c\epsilon_b\frac{\gamma^2}{8 \pi}\frac{\sqrt{\frac{\delta_c}{2}|O_2|\cosh\left(4 G \gamma P_2\right)}}{\frac{1}{\delta_b}\sqrt{1 - b_o^2 \tanh^2\left[G \gamma b_o\tilde T\right]}}.
\end{equation}

On shell, we have $O_1=-O_2$. The reduced action can be easily computed 
\begin{equation}\label{eq:act-P2-lqc}
S_{\rm red} = \int d\tilde T \left(P_2\dot{O}_2-\tilde H_{\rm red}\right),
\end{equation}
where $\tilde H_{\rm red}=-O_2$ is the reduced Hamiltonian. The dynamics for this gauge fixing involves that $O_2$ is a constant of the motion and $P_2=T+P_2^0$, with $P_2^0$ another constant of the motion. 

As we did in Sec. \ref{sec:eff-dyn}, we will start with the identification of the Dirac observables $o_2$ and $O_2$ in classical GR and effective LQC, respectively. This amount to
\begin{equation}
O_2= 2GM\gamma L_o.
\end{equation}

As we already considered there, we determine the time $\tilde T_{\rm hor}$ and the observable $P_2^0$ by means of the conditions on the connections $c(\tilde T_{\rm hor})=\gamma L_o/(4GM)$ and $b(\tilde T_{\rm hor})=0$. This involves
\begin{equation}
P_2^0=\frac{1}{4\gamma G}\log\left(\frac{8GM}{\gamma L_o\delta_c}\right)-\tilde T_{\rm hor},
\quad \tilde T_{\rm hor} = \frac{1}{b_o G\gamma}\arctanh\left[\frac{1}{b_o}\right].
\end{equation}

In these circumstances, the original phase space variables are given by
\begin{align}\label{eq:cbT-lqc}
c(\tilde T) &= \frac{2}{\delta_c}\arctan\left(e^{-4 G \gamma  (\tilde T+P_2^0)}\right),\quad p_c(\tilde T)=\epsilon_c\frac{\delta_c}{2}O_2\cosh\left[4 G \gamma \left(\tilde T+P_2^0\right)\right], \\
b(\tilde T) &= -\epsilon_b\frac{1}{\delta_b}\arccos\left[b_o\tanh\left(G \gamma b_o \tilde T\right)\right], \\
p_b(\tilde T)&=-\epsilon_b\frac{\delta_b O_2}{b_o^2} \cosh^2\left(G \gamma \tilde T b_o\right) \sqrt{1 - b_o^2 \tanh^2\left[G \gamma \tilde T b_o\right]}.
\end{align}

\end{document}